\documentclass[journal,comsoc]{IEEEtran}

\usepackage{xspace,amsmath,amssymb,amsfonts,epsfig,syntonly}
\usepackage{cite,bm,color,url,textcomp,empheq,boxedminipage}
\usepackage{algorithmicx,algorithm,algpseudocode}
\usepackage{epstopdf}
\usepackage{empheq}
\usepackage{pifont}
\usepackage{setspace}

\usepackage{graphicx,graphics,framed,subfigure}  
\usepackage{multirow,multicol}
\usepackage{psfrag}    
\usepackage{stfloats}
\usepackage{url}

\newtheorem{assumption}{Assumption}
\newtheorem{theorem}{{Theorem}}
\newtheorem{lemma}{{Lemma}}

\newtheorem{proposition}{{Proposition}}
\newtheorem{example}{{Example}}
\newtheorem{remark}{{Remark}}

\DeclareMathOperator*{\argmax}{arg\,max}
\DeclareMathOperator*{\argmin}{arg\,min}

\long\def\symbolfootnote[#1]#2{\begingroup
\def\thefootnote{\fnsymbol{footnote}}
\footnote[#1]{#2}\endgroup}

\psfull

\allowdisplaybreaks[4]

\usepackage[T1]{fontenc}
\usepackage{ifpdf}
 \ifpdf
 \else
 \fi

\begin{document}
\title{Real-Time Resource Allocation for Wireless Powered Multiuser Mobile Edge Computing With Energy and Task Causality}

\author{Feng Wang,~\IEEEmembership{Member, IEEE}, Hong Xing,~\IEEEmembership{Member, IEEE}, and Jie Xu,~\IEEEmembership{Member, IEEE}

\thanks{This paper was presented in part at the IEEE International Communications Conference (ICC), Shanghai, China, May 20--24, 2019~\cite{Conf_version}.} 

\thanks{F. Wang is with the School of Information Engineering, Guangdong University of Technology, Guangzhou 510006, China (e-mail: fengwang13@gdut.edu.cn).}

\thanks{H. Xing is with the College of Information Engineering, Shenzhen University, Shenzhen 518060, China, and also with the Department of Engineering, King's College London, London WC2R 2LS, U.K. (e-mail: hong.xing@szu.edu.cn).}

\thanks{J. Xu is with the Future Network of Intelligence Institute (FNii) and the School of Science and Engineering, the Chinese University of Hong Kong (Shenzhen), Shenzhen 518172, China (e-mail: xujie@cuhk.edu.cn). J. Xu is the corresponding author.}
}

\maketitle

\begin{abstract}
  This paper considers a wireless powered multiuser mobile edge computing (MEC) system, in which a multi-antenna hybrid access point (AP) wirelessly charges multiple users, and each user relies on the harvested energy to execute computation tasks. We jointly optimize the energy beamforming and remote task execution at the AP, as well as the local computing and task offloading, aiming to minimize the total system energy consumption over a finite time horizon, subject to causality constraints for both energy harvesting and task arrival at the users. In particular, we consider a practical scenario with casual task state information (TSI) and channel state information (CSI), i.e., only the current and previous TSI and CSI are available, but the future TSI and CSI can only be predicted subject to certain errors. To solve this real-time resource allocation problem, we propose an offline-optimization inspired online design approach. First, we consider the offline optimization case by assuming that the TSI and CSI are perfectly known {\em a-priori}. In this case, the energy minimization problem corresponds to a convex problem, for which the semi-closed-form optimal solution is obtained via the Lagrange duality method. Next, inspired by the optimal offline solution, we propose a sliding-window based online resource allocation design in practical cases by integrating with the sequential optimization. Finally, numerical results show that the proposed joint wireless powered MEC designs significantly improve the system's energy efficiency, as compared with the benchmark schemes that consider a sliding window of size one or without such joint optimization.
\end{abstract}


\section{Introduction}

Currently, the integration of mobile edge computing (MEC)\cite{Sar14,Qi19,JunZhang17,Xiaowen18,Hong19,TCOM19,Xing19} and wireless power transfer (WPT)\cite{Zeng17,Clerckx,Rui13,Xu14,Xie19}, namely {\em wireless powered MEC}, has emerged as a new paradigm to achieve self-sustainable mobile computing for latency-sensitive and computation-intensive applications\cite{You16,Feng17,Bi18}. In typical wireless powered MEC systems, hybrid access points (APs), integrated with MEC servers and energy transmitters (ETs), are deployed to wirelessly charge a large number of low-power wireless devices (such as sensors and wearable devices), and these devices rely on the harvested energy to execute their computation tasks via local computing by themselves and/or task offloading to the AP for remote task execution therein. Thanks to the controllability of the WPT, the interaction between the users' computation energy demand and the APs' wireless power supply can be effectively adjusted and balanced in wireless powered MEC systems. It is thus envisioned that the proposed wireless powered MEC solution can significantly enhance the computation performance and energy-sustainability in future Internet-of-things (IoT) sensing/monitoring applications~\cite{JunZhang17}.

Unlike the conventional MEC with fixed battery-powered devices\cite{Sar14,Qi19,JunZhang17,Xing19,Xiaowen18,TCOM19} and wireless powered communications\cite{Zeng17,Clerckx,Rui13}, the design of wireless powered MEC systems encounters several new technical challenges due to the complicated interplay among the WPT, computation, and communication. First, wireless devices powered by WPT are subject to the energy harvesting constraints, such that the amount of energy harvested by each user from the WPT cannot be larger than that of energy consumed for computation. It thus calls for a joint management of the harvested energy, computation, and communication resource allocations, in order to achieve an energy supply-demand balance. Second, multiple devices need to share the limited energy, communication, and computation resources with each other. There thus exists another fundamental tradeoff between the system computation performance and the user fairness. Third, due to the bursty nature of computation traffics, the task arrival amounts at users may fluctuate substantially over time, whilst both the WPT and computation offloading performances are subject to their wireless channel fading conditions between the users and APs. For instance, if the WPT or offloading channels stay in deep fading or the computation tasks are not timely offloaded, then the corresponding devices may face energy deficit and cannot meet the computation requirement in due course. Furthermore, the task state information (TSI) (i.e., task arrival timing and amount) and channel state information (CSI) are only known causally at the AP and users; i.e., at any given time instance, the AP and the users can only obtain the current and previous TSI/CSI, but the future TSI/CSI predicated with certain errors. It is thus crucial to adapt task allocation to task arrivals and task offloading dynamically.

\subsection{Related Work}

For wireless powered MEC system designs, a fundamental question to be addressed is as follows: taking into account the task arrival causality and energy harvesting constraints with causal TSI and CSI, how to jointly design the WPT, computation, and communication resource allocations so as to minimize the wireless powered MEC system's total energy consumption? In the literature, there have been several prior works investigating joint WPT, communication, and computation resource allocations in wireless powered MEC systems under various setups e.g., one single user \cite{You16}, multiple users \cite{Feng17,Bi18}, and user cooperation \cite{Hu18,Wu18,He18}. Note that these works \cite{You16,Feng17,Bi18,Hu18,Wu18,He18} focused on one-shot optimization at a particular time slot by assuming unchanged wireless channels and static task models at users, which cannot address the new design challenges due to the dynamic task arrivals in practical systems. The work in \cite{Zhou18} jointly optimized the task offloading and trajectory for unmanned aerial vehicle (UAV)-enabled MEC systems, where the UAVs are equipped with MEC servers. Under the binary offloading policy, the work in \cite{Bi19} studied the deep reinforcement algorithms to optimize the system computing performance. In a different MEC context by considering renewable energy harvesting devices and dynamic task arrivals over time, the recent works \cite{Mao16JASC,Xiao19,Jie17Cog,Wei19} studied joint communication and computation managements, where the sophisticated Lyapunov optimization tools are leveraged to design resource management policies for long-term system cost minimization.

Very recently, based on a harvest-then-transmit protocol, \cite{JieFeng19} considered the computation outage performance for offloading and studied the wireless powered MEC design in a delay-limited case when all tasks are expected to be executed by the end of a time slot. Considering dynamic task arrivals over multiple time slots, \cite{Feng19} investigated the optimal offline joint-WPT-MEC design structure for a basic wireless powered {\em single-user} MEC system with separately located AP/ET. Nevertheless, there still lacks a joint energy-efficient design of the WPT at the AP and the communication/computation energy demand at the users for wireless powered MEC systems, by taking into account both energy and task dynamics (and thus causality) over time, which thus motivates our study in this work.

\subsection{Contributions}
In this paper, we investigate the wireless powered multiuser MEC designs with time-varying task arrivals, where a multi-antenna hybrid AP employs the energy beamforming to wirelessly charge the users over the air, and each user utilizes the harvested energy to execute its latency-constrained tasks. It is assumed that the downlink WPT from the AP to the users and the uplink task offloading are operated simultaneously over orthogonal frequency bands. In order to avoid interference among the users, we assume that the frequency-division multiple access (FDMA) protocol is employed to enable simultaneous task offloading from multiple users to the AP. The users need to successfully complete their computation tasks by local computing and/or task offloading within a finite time horizon consisting of multiple equal-duration time slots. The main results of this paper are summarized as follows.
\begin{itemize}
\item
  First, we model the TSI for MEC, the CSI for WPT, and the CSI for task offloading with predicable and additive errors following arbitrary distribution. Under such considerations, we minimize the total system energy consumption (including AP's transmission energy and remote task execution energy) over the whole horizon subject to individual energy harvesting and task arrival causality constraints at the users, by jointly optimizing energy beamforming and remote task execution at the AP, as well as the local computing and task offloading strategies at the users over different slots.
\item
  Next, we propose an offine-inspired-online design (i.e., sliding-window based) scheme to solve the latency-constrained energy minimization problem in real time. By assuming the TSI/CSI are perfectly known {\em a-priori}, we first derive the optimal offline solution based on the Lagrange duality method, and reveal the optimal (i.e., monotonic) structures of task allocation for the users' local computing and the AP's remote execution, respectively. Then, inspired by the optimal offline solution, we develop an online sliding-window based scheme with causally known TSI/CSI, by further utilizing a ``sliding-window'' based sequential optimization approach.
\item
  Finally, numerical results are provided to demonstrate the energy efficiency of the optimal offline and online joint-WPT-MEC designs, as compared with the benchmark schemes without such joint design or with a short-sighted ({\em myopic}) based design that assumes complete task allocation of the present task arrival. We also show the robustness of the proposed online design scheme against the predicted errors of TSI/CSI.
\end{itemize}

The remainder of this paper is organized as follows. Section II presents the wireless powered multiuser MEC system model. Section III formulates the system's total energy minimization problem over a finite time horizon. Section IV presents the optimal offline solution to the formulated problem. Inspired by the offline optimal solution, Section V proposes an online sliding-window based scheme to solve the formulated problem in real time. Section VI provides numerical results to evaluate the proposed designs as compared to other benchmark schemes. Finally, Section VII concludes this paper.

{\em Notations}: Throughout this paper, boldface lowercase (uppercase) letters indicate vectors (matrices). We use $\mathbb{R}^{N\times M}$ and $\mathbb{C}^{N\times M}$ to denote the set of real-valued and complex-valued $N\times M$ matrices, respectively; $\bm I_N$ denotes an $N\times N$ identity matrix; $\bm 0$ denotes the vector or matrix with an appropriate dimension. For a matrix $\bm X$, we use ${\rm tr}(\bm X)$, $\bm X^\dagger$, and $\bm X^H$ to denote its trace, transposition, and conjugate transposition, respectively. For a symmetric matrix $\bm S$, $\bm S \succeq \bm 0$ indicates that $\bm S$ is positive semidefinite. For a vector $\bm x$, its Euclidean norm is denoted by $\|\bm x\|$. The distribution of a circularly symmetric complex Gaussian (CSCG) vector $\bm x$ with mean vector $\bar{\bm x}$ and covariance matrix $\bm \Sigma$ is denoted as $\bm x \sim {\cal CN}(\bar{\bm x},\bm \Sigma)$, and the uniform distribution of a real-valued $x$ within the interval $[a,b]$ is denoted as $x\sim{\cal U}[a,b]$, where $\sim$ stands for ``distributed as''. Finally, $[x]^+$ denotes $\max(x,0)$ and $\mathbb{E}[\cdot]$ represents the expectation operation.

\section{System Model}\label{Sec:System}

\begin{figure}
  \centering
  \includegraphics[width = 3.5in]{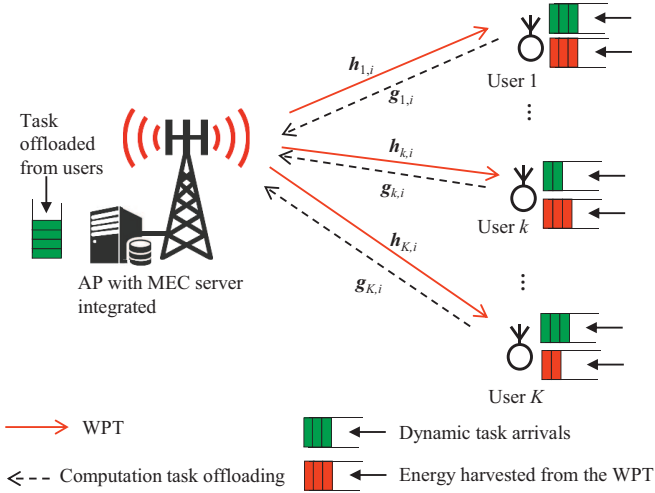}
 \caption{The wireless powered multiuser MEC system model.} \label{fig.system-model}
\end{figure}

We consider a wireless powered multiuser MEC system as shown in Fig.~\ref{fig.system-model}, where an $N_t$-antenna hybrid AP employs the energy beamforming to simultaneously charge a set ${\cal K}\triangleq \{1,\ldots,K\}$ of single-antenna users over the air, and each user relies on the harvested wireless energy to execute its latency-constrained computation tasks. In particular, we consider a finite time horizon with duration $T$, which is divided into $N$ time slots each with duration $\tau=T/N$. Let ${\cal N}\triangleq \{1,\ldots,N\}$ denote the set of the $N$ slots. For each user $k\in{\cal K}$, the computation tasks are assumed to arrive at the beginning of each slot $i\in{\cal N}$. In order to access the AP's remote task execution, these $K$ users need to offload their computation tasks to the AP. After the MEC server successfully executes the offloaded tasks, the AP sends the computation results back to respective users. It is assumed that the $K$ users are subject to a {\em common} completion deadline of $T$, i.e., each user needs to successfully execute its respective tasks by the end of this time horizon, which is motivated by some IoT sensing/monitoring applications\cite{JunZhang17,Savazzi20}. For example, wireless users may need to cooperatively perform some computation tasks (e.g., for data analysis) after certain amount of data sensing (e.g., temperature, humidity, pressure, tactility) and pre-processing at each individual user are completed. In addition, since the size of computation results is generally much smaller than that of task input-bits, this paper focuses on the time and energy consumed by task offloading from the users to the AP in the uplink, and ignores those consumed by the computation result downloading in the downlink.

For the purpose of idea exposition, in this paper we consider a flat-fading environment, in which the fading channels for WPT and task offloading remain unchanged within one time slot but they may change slot by slot. Furthermore, as widely adopted in the MEC literature\cite{Sar14,Qi19,JunZhang17,Feng17,You16,Bi18}, in this paper we focus on independent task models, in which the task arrivals of one user and those among different users are all independent of each other. We denote $\bm h_{k,i}\in \mathbb{C}^{N_t\times 1}$ and $\bm g_{k,i}\in\mathbb{C}^{N_t\times 1}$ as the channel vectors for the WPT from the AP to user $k$ and for the task offloading from user $k$ to the AP at slot $i\in{\cal N}$, respectively. We denote $A_{k,i}\geq 0$ as the number of task input-bits arriving at user $k$ during slot $i$. Throughout the paper, we make the following assumption on modelling the predicable TSI/CSI in wireless powered MEC systems, which is reasonably valid in practice.

\begin{assumption}[Predicable TSI/CSI with Additive Error]
The TSI for MEC, the CSI for WPT, and the CSI for task offloading over the horizon (i.e., $\{ A_{k,i},\bm h_{k,i},\bm g_{k,i}\}$) are predictable but with additive errors. The TSI/CSI predications can be profiled based on the historical data records and computation patterns in targeted applications by, e.g., machine learning algorithms\cite{Jie17Cog,Savazzi20,Bi19}. By denoting $\hat{A}_{k,i}\in
\mathbb{R}$ ($\hat{\bm h}_{k,i}\in\mathbb{C}^{N_t\times1}$, $\hat{\bm g}_{k,i}\in\mathbb{C}^{N_t\times1}$) and $\delta_{A_{k,i}}\in
\mathbb{R}$ ($\bm \delta_{h_{k,i}}\in\mathbb{C}^{N_t\times1}$, $\bm \delta_{g_{k,i}}\in\mathbb{C}^{N_t\times1}$) as the predicated TSI (CSI) and the additive TSI (CSI) error for user $k$ at slot $i$, respectively, we establish the following additive TSI/CSI error model:
\begin{align}\label{eq.prediction}
&A_{k,i} = \hat{A}_{k,i} + \delta_{A_{k,i}} \notag \\
&\bm h_{k,i} = \hat{\bm h}_{k,i} + {\bm \delta}_{h_{k,i}} \notag \\
&\bm g_{k,i} = \hat{\bm g}_{k,i} + {\bm \delta}_{g_{k,i}},
\end{align}
where $k\in{\cal K}$ and $i\in{\cal N}$.
\end{assumption}

In the following, we first introduce the task execution at the users via local computing and task offloading, and then present the WPT and remote task execution at the AP.

\subsection{Local Computing and Task Offloading at Users}
During each time slot $i\in{\cal N}$, each user $k\in{\cal K}$ needs to execute its tasks via local computing by itself and/or offloading for the AP's remote execution. By assuming that the computation tasks of $A_{k,i}$ input-bits at slot $i$ is partitionable\cite{JunZhang17}, user $k$ can employ the partial offloading technique to arbitrarily partition its computation tasks into two parts for local computing and offloading, respectively. Let $l^{\rm loc}_{k,i}\geq 0$ and $l^{\rm off}_{k,i}\geq 0$ denote the number of task input-bits for local computing and offloading of user $k\in{\cal K}$ at slot $i$, respectively. Since the cumulative number of task input-bits executed until slot $i\in{\cal N}\setminus\{N\}$ (i.e., $\sum_{j=1}^i l^{\rm loc}_{k,j}+\sum_{j=1}^i l^{\rm off}_{k,j}$) cannot exceed that having already arrived at user $k$ (i.e., $\sum_{j=1}^i A_{k,j}$), we establish the {\em task causality constraints} at the users, which are expressed as
\begin{align}\label{eq.Si-user}
&\sum_{j=1}^i A_{k,j} -\sum_{j=1}^i l^{\rm loc}_{k,j}-\sum_{j=1}^i l^{\rm off}_{k,j} \geq 0,~\forall i\in{\cal N}\setminus\{N\}, ~k \in{\cal K}.
\end{align}
In addition, due to the common computation latency requirement, each user $k\in{\cal K}$ needs to accomplish the task execution by the end of the last time slot $N$. Therefore, we impose the {\em computation deadline constraints} at the users as
\begin{align}\label{eq.SN-user}
\sum_{j=1}^{N} A_{k,j} - \sum_{j=1}^N l^{\rm loc}_{k,j} - \sum_{j=1}^{N-1} l^{\rm off}_{k,j} = 0, ~~\forall k\in {\cal K}.
\end{align}

In the following, we introduce the local computing and task offloading for each user $k\in{\cal K}$, respectively.

\subsubsection{Local Computing at Users}
First, with regards to local computing of the $l^{\rm loc}_{k,i}$ task input-bits at user $k\in{\cal K}$ at time slot $i\in{\cal N}$, a total amount of $C_kl_{k,i}^{\rm loc}$ central processing unit (CPU) cycles is required to run, where $C_k\geq 0$ (in CPU cycles per bit) denotes the number of CPU cycles required for executing one task input-bit at user $k\in{\cal K}$. In order to maximize the energy efficiency for local computing, each user $k\in{\cal K}$ can apply the dynamic voltage and frequency scaling (DVFS) technique by adjusting the CPU frequency as $C_kl^{\rm loc}_{k,i}/\tau$ (in CPU cycles per second) during each slot $i\in{\cal N}$\cite{Feng17}. As a result, the energy consumption $E_{k,i}^{\rm loc}$ for local computing at user $k\in{\cal K}$ at slot $i\in{\cal N}$ is given as
\begin{align}\label{eq.Eki}
E_{k,i}^{\rm loc} = \sum_{n=1}^{C_k l^{\rm loc}_{k,i}} \zeta_k \left(\frac{C_k l^{\rm loc}_{k,i}}{\tau} \right)^2 =\frac{\zeta_k C_k^3 (l^{\rm loc}_{k,i})^3}{\tau^2},
\end{align}
where $\zeta_{k}> 0$ is the effective switched capacitance coefficient decided by user $k$'s CPU chip architecture\cite{You16,Feng17,Bi18}.

\subsubsection{Task Offloading from Users to AP}
Next, with regards to task offloading of the $l^{\rm off}_{k,i}$ task input-bits to the AP, we adopt a FDMA protocol for the $K$ users to simultaneously offload their tasks to the AP, where each user $k\in{\cal K}$ is allocated with an equal bandwidth of $B>0$ (Hz). Under the assumption of the maximal ratio combining (MRC) receiver adopted at the AP, the offloading rate is expressed as $r_{k,i}=B\log_2(1+p_{k,i}\|\bm g_{k,i}\|^2/{\Gamma \sigma^2})$, where $p_{k,i}$ denotes the user $k$'s transmission power at time slot $i$, $\sigma^2$ is the additive white Gaussian noise (AWGN) power at the AP's receiver, and $\Gamma\geq1$ denotes the signal-to-noise ratio (SNR) penalty due to the practical modulation and coding scheme employed\cite{Goldsmith}. For the conciseness of notation, we set $\Gamma=1$ in the sequel. Accordingly, we have $r_{k,i}=l^{\rm off}_{k,i}/\tau$, and user $k$'s energy consumption on computation offloading at slot $i\in{\cal N}$ of duration $\tau$ is expressed as
\begin{align}
E_{k,i}^{\rm off} = p_i\tau = \frac{\tau\sigma^2(2^{\frac{l^{\rm off}_{k,i}}{\tau B}}-1)}{\|\bm g_{k,i}\|^2}.
\end{align}

\subsection{WPT and Remote Task Execution at AP}
In the wireless powered MEC system, in addition of transceiving information, the AP also needs to wirelessly charge the $K$ users by energy beamforming and remotely execute the offloaded tasks from the $K$ users. In the following, we introduce the WPT from the AP to the users and the remote task execution at the AP, respectively.

\subsubsection{WPT from AP to Users}
First, we consider the energy beamforming for WPT from the AP to the $K$ users at each slot $i\in{\cal N}$. Let $\bm s_i\in\mathbb{C}^{N_t\times 1}$ denote the energy-bearing transmit signal and $\bm S_i\triangleq \mathbb{E}[\bm s_i\bm s_i^H]\in\mathbb{C}^{N_t\times N_t}$ denote the transmit covariance matrix of the AP at slot $i\in{\cal N }$. For ease of analysis, we consider that the input RF power of each user $k\in{\cal K}$ is always within the linear regime of the rectifier. Hence, adopting the widely used linear energy harvesting model\cite{Zeng17,Clerckx,Rui13}, the amount of energy harvested by user $k\in{\cal K}$ at slot $i\in{\cal N}$ is given as
\begin{align}\label{eq.EH-model}
E_{k,i}^{\rm har}=\tau \eta_k \mathbb{E}\left[|\bm s^H_i \bm h_{k,i}|^2\right]=\tau \eta_k {\rm tr}\left(\bm S_i {\bm h}_{k,i}{\bm h}_{k,i}^H \right),
\end{align}
where $\eta_k\in(0,1]$ denotes the energy harvesting efficiency of user $k$, the expectation $\mathbb{E}[\cdot]$ is taken on the random energy-bearing signal $\bm s_i$, and ${\rm tr}(\cdot)$ represents the trace operation for matrices.

\subsubsection{Remote Task Execution at AP}
Next, we consider the remote task execution at the AP. We denote $l^{\rm mec}_{0,i}\geq 0$ as the number of task input-bits executed the AP at slot $i$. Due to the task causality, the earliest the tasks offloaded from the users at time slot $i$ can be executed at the AP is time slot $(i+1)$. Therefore, at each time slot $i\in{\cal N}\setminus\{1\}$, the number of task input-bits cumulatively executed by AP (i.e., $\sum_{j=2}^i l^{\rm mec}_{0,j}$) cannot exceed that cumulatively offloaded from all the $K$ users up to the previous slots $\{1,\ldots,i-1\}$ (i.e., $\sum_{j=1}^{i-1} \sum_{k=1}^K l^{\rm off}_{k,j}$). As in \eqref{eq.Si-user}, we establish the task causality constraints at the AP, which are expressed as
\begin{align}\label{eq.MECi}
\sum_{j=1}^{i-1} \sum_{k=1}^K l^{\rm off}_{k,j} - \sum_{j=2}^i l^{\rm mec}_{0,j} \geq 0,~~\forall i\in \cal{N}\setminus\{N\}.
\end{align}

In addition, due to the common computation latency requirement, the AP needs to complete all the task execution before the end of the last slot $N$. As in \eqref{eq.SN-user}, we express the computation deadline constraint for the AP as
\begin{align}\label{eq.MECN}
\sum_{j=1}^{N-1} \sum_{k=1}^K l^{\rm off}_{k,j} - \sum_{j=2}^N l^{\rm mec}_{0,j} = 0.
\end{align}
It is worth noting that $l_{0,1}^{\rm mec}=0$ and $l^{\rm off}_{k,N}=0$, $\forall k\in{\cal K}$, since there is no offloaded task awaiting execution at the AP at the first slot, and there is no time left for the AP to perform remote task execution at the last time slot, respectively.

Furthermore, let $C_0\geq 0$ denote the number of required CPU cycles for executing one task input-bit at the AP. Analogously to local computing at the $K$ users, the AP employs the DVFS technique to adjust its MEC server's CPU frequency as $C_0 l^{\rm mec}_{0,i}/\tau$ (in CPU cycles per second) at slot $i$, such that the computation energy is minimized at the AP\cite{Feng17}. Therefore, the total amount of computation energy consumed over the $N$-slot horizon is expressed as $\sum_{i=2}^N \frac{\zeta_0 C_0^3 (l^{\rm mec}_{0,i})^3}{\tau^2}$, where $\zeta_0$ is the capacitance coefficient specified by the MEC server's CPU architecture at the AP.

Note that each user $k\in{\cal K}$ relies on the energy beamforming of the AP to implement its local computing and task offloading simultaneously. By assuming a sufficiently large energy buffer employed at each user, the wireless energy harvested at time slot $i$ of the horizon under consideration can be utilized at time slot $i$. In this case, at each time slot $i\in{\cal N}$, the cumulatively consumed energy at each user (i.e., $\sum_{j=1}^i E_{k,j}^{\rm loc}+\sum_{j=1}^i E_{k,j}^{\rm off}$) until that time slot cannot exceed that cumulatively harvested from the AP up to the present time slot (i.e., $\sum_{j=1}^{i} E^{\rm har}_{k,j}$); in other words, it holds that $\sum_{j=1}^i E_{k,j}^{\rm loc}+\sum_{j=1}^i E_{k,j}^{\rm off} \leq \sum_{j=1}^{i} E^{\rm har}_{k,j}$, $\forall i\in{\cal N}, k\in{\cal K}$. As a result, we establish the individual {\em energy harvesting constraints} for the users\cite{Feng19,Rui12,Li15,XunZhou16}, which are expressed as
\begin{align}\label{eq.EHi}
&\sum_{j=1}^i\Big( \frac{\zeta_k C_k^3 (l^{\rm loc}_{k,j})^3}{\tau^2} + \frac{\tau\sigma^2}{\|{\bm g}_{k,j}\|^2}(2^{\frac{l^{\rm off}_{k,j}}{\tau B}}-1)\Big) \notag \\
& \quad\quad\quad\quad \leq  \sum_{j=1}^i \tau \eta_k {\rm tr}\left(\bm S_j\bm h_{k,j}\bm h_{k,j}^H \right), ~~\forall i\in {\cal N},~k\in{\cal K}.
\end{align}

\section{Problem Formulation}
In this paper, we pursue an energy-efficient joint-WPT-MEC design by minimizing the system's total energy consumption over the horizon subject to the task causality and computation deadline constraints at the AP and users, as well as the energy harvesting constraints. Specifically, the design objective is sum of the AP's transmission energy for WPT and remote task execution energy, i.e., $\sum_{i=1}^N\tau{\rm tr}(\bm S_i) + \sum_{i=2}^N\frac{\zeta_0 C_0^3 (l^{\rm mec}_{0,i})^3}{\tau^2}$. The decision variables include the AP's transmit covariance matrices $\{\bm S_i\}$ for WPT, the number of task input-bits $\{l^{\rm mec}_{0,i}\}$ for remote task execution at the AP, and the number of task input-bits $\{l^{\rm loc}_{k,i}\}$ and $\{l^{\rm off}_{k,i}\}$ for users' local computing and offloading. Mathematically, the system's total energy minimization problem is formulated as
\begin{subequations}\label{eq.prob1}
\begin{align}
&({\rm P}1):\notag \\
&\min_{\{\bm S_i,l^{\rm mec}_{0,i},l^{\rm loc}_{k,i},l^{\rm off}_{k,i}\}} \sum_{i=1}^{N}\tau{\rm tr}(\bm S_i) + \sum_{i=2}^N \frac{\zeta_0 C_0^3 (l^{\rm mec}_{0,i})^3}{\tau^2}  \\
&\quad\quad {\rm s.t.}~ \sum_{j=1}^i\Big( \frac{\zeta_k C_k^3 (l^{\rm loc}_{k,j})^3}{\tau^2} + \frac{\tau\sigma^2}{\|{\bm g}_{k,j}\|^2}(2^{\frac{l^{\rm off}_{k,j}}{\tau B}}-1)\Big) \notag \\
&~~\quad\quad\quad \leq  \sum_{j=1}^i \tau \eta_k {\rm tr}\left(\bm S_j\bm h_{k,j}\bm h_{k,j}^H \right),~\forall i\in {\cal N},k \in{\cal K}\\
&\quad\quad\quad\sum_{j=1}^i A_{k,j} -\sum_{j=1}^i l^{\rm loc}_{k,j}-\sum_{j=1}^i l^{\rm off}_{k,j} \geq 0,\notag \\
&\quad\quad\quad\quad\quad\quad\quad\quad\quad\quad\quad\quad \forall i\in {\cal N}\setminus\{N\}, ~k \in{\cal K}\\
&\quad\quad\quad\sum_{j=1}^{N} A_{k,j} - \sum_{j=1}^N l^{\rm loc}_{k,j} - \sum_{j=1}^{N-1} l^{\rm off}_{k,j} = 0, ~\forall k\in {\cal K}\\
&\quad\quad\quad \sum_{j=1}^{i-1} \sum_{k=1}^K l^{\rm off}_{k,j} - \sum_{j=2}^i l^{\rm mec}_{0,j} \geq 0,~~\forall i\in \cal{N}\setminus\{N\}\\
&\quad\quad\quad \sum_{j=1}^{N-1} \sum_{k=1}^K l^{\rm off}_{k,j} - \sum_{j=2}^N l^{\rm mec}_{0,j} = 0\\
&\quad\quad\quad \bm S_i\succeq \bm 0, l^{\rm loc}_{k,i}\geq 0,~\forall i\in{\cal N}, k\in{\cal K} \\
&\quad\quad\quad l^{\rm mec}_{0,j}\geq 0, \forall j\in{\cal N}\setminus\{1\}, l^{\rm off}_{k,i}\geq 0,\notag \\
&\quad\quad\quad\quad\quad\quad\quad\quad\quad\quad\quad\quad\forall i\in{\cal N}\setminus\{N\},k\in{\cal K}. \label{eq.prime-fea}
\end{align}
\end{subequations}
In problem (P1), (\ref{eq.prob1}b) denotes the users' energy harvesting constraints over time slots; (\ref{eq.prob1}c) and (\ref{eq.prob1}d) denote the users' task causality and computation deadline constraints, respectively; and (\ref{eq.prob1}e) and (\ref{eq.prob1}f) denote the causality and the computation deadline constraints for the AP's remote task execution, respectively.

\begin{remark}
In practice, at each time slot $i$, the current and previous TSI/CSI $\{A_{k,j},\bm h_{k,j},\bm g_{k,j}\}_{j=1}^i$ can be perfectly known, but the future TSI/CSI $\{A_{k,j},\bm h_{k,j},\bm g_{k,j}\}_{j=i+1}^N$ are only partially predicated with additive errors $\{A_{k,j},\hat{\bm h}_{k,j},\hat{\bm g}_{k,j}\}_{j=i+1}^N$. In other words, the knowledge $\{A_{k,j},\bm h_{k,j},\bm g_{k,j}\}$ can only be causally obtained by the AP and users. Therefore, problem (P1) is a very challenging optimization problem to be optimally solved. One commonly used method to solve an online optimization problem like (P1) is through the dynamic programming (DP) based methods\cite{DP}, which could provide the optimal online solution when $\{\delta_{A_{k,j}}, \bm \delta_{h_{k,j}}, \bm \delta_{g_{k,j}}\}$ are modelled as stochastic processes with known distributions. However, due to the ``curse of dimensionality'' issue and the fact that TSI/CSI predication might not be modeled as stochastic processes with known distributions, the optimal DP-based approaches are computationally prohibitive to solve the online problem of (P1). To our best knowledge, there still lack practical solutions for handling such multiuser joint-WPT-MEC designs with TSI/CSI uncertainties over time.
\end{remark}

Along with gaining some clear joint-WPT-MEC deign insights, in this paper we propose a novel {\em offline-inspired-online} approach to solve the challenging online problem (P1). In Section IV, we first consider the offline optimization case with the TSI/CSI perfectly known {\em a-priori}, which serves as a performance upper bound for any online designs of wireless powered MEC systems and also provides some insights for them. In the offline case, it always holds that $\delta_{A_{k,i}}=0$, $\bm \delta_{h_{k,i}}=\bm 0$, and $\bm \delta_{g_{k,i}}=\bm 0$, $\forall i\in{\cal N}$, $k\in{\cal K}$, and therefore problem (P1) is a convex optimization problem\cite{Boyd2004}. As such, leveraging the Lagrange duality method, in Section IV we obtain the well-structured optimal solution to the offline problem (P1). Inspired by this optimal offline solution and integrating a sequential optimization approach\cite{Rahbar15}, in Section V we further develop the online sliding-window based scheme to effectively solve problem (P1) in real time, which can handle the cases with arbitrary/unknown distributions of TSI/CSI predication errors, even when the user number and the slot number both grow large.

\section{Optimal Offline Solution to Problem (P1)}\label{sec:optimal}
In this section, we consider the offline optimization of problem (P1) in the sense that the TSI/CSI ($\{A_{k,i},\bm h_{k,i},\bm g_{k,i}\}_{j=1}^N$) are perfectly known {\em a-priori}. We first present the optimal offline solution to problem (P1), and then we reveal the essential insights for energy-efficient multiuser joint-WPT-MEC designs with dynamic task arrivals.

To this end, we establish the lemma on the convexity of problem (P1) in the offline case.

\begin{lemma}\label{lem.offline}
In the offline case, problem (P1) is a convex optimization problem.
\end{lemma}
\begin{IEEEproof}
First, since function ${\rm tr}(\bm S_i)$ is affine with respect to the positive semidefinite matrix $\bm S_i\succeq \bm 0$ and function $x^3$ is convex with respect to $x\geq 0$\cite{Boyd2004}, the objective function of (P1) is convex with respect to the optimization variables. Next, in the offline case, since the TSI/CSI (i.e., $\{A_{k,i},\bm h_{k,i},\bm g_{k,i}\}_{i=1}^N$) are perfectly known {\em a-priori}, the left-hand-side (LHS) and the right-hand-side (RHS) of (\ref{eq.prob1}b) become convex and linear functions, respectively. Hence, (\ref{eq.prob1}b) defines a convex set of $\{\bm S_i,l^{\rm mec}_{0,i},l^{\rm loc}_{k,i},l^{\rm off}_{k,i}\}$. Likewise, since the LHSs of (\ref{eq.prob1}c)--(\ref{eq.prob1}h) are affine functions, each of (\ref{eq.prob1}c)--(\ref{eq.prob1}h) defines a convex set\cite{Boyd2004}. Therefore, problem (P1) in the offline case is a convex optimization problem.
\end{IEEEproof}

Based on Lemma~\ref{lem.offline}, since the offline problem (P1) satisfies the Slater's condition, the strong duality holds between (P1) and its dual problem\cite{Boyd2004}. Based on the Lagrange duality method, we can thus derive the optimal offline solution to problem (P1), which is denoted as $\{\bm S^{*}_{i}, l^{\rm mec *}_{0,i}, l^{\rm loc *}_{k,i}, l^{\rm off *}_{k,i}\}$. Formally, we establish the following theorem.

\begin{theorem} \label{thorem1}
The optimal offline solution $\{\bm S^{*}_{i}, l^{\rm mec *}_{0,i},l^{\rm loc *}_{k,i},l^{\rm off *}_{k,i}\}$ to problem (P1) is given by
\begin{subequations}\label{eq.optimal-sol}
\begin{align}
 & l^{\rm mec *}_{0,i} = \frac{\tau}{\sqrt{3}\zeta^{\frac{1}{2}}_0C_0^{\frac{3}{2}}}\sqrt{\big(\textstyle -\sum_{j=i}^N\nu^*_j\big)^+}, ~~\forall i\in{\cal N}\setminus\{1\} \\
 & l_{k,i}^{\rm loc *} = \frac{\tau}{\sqrt{3}\zeta_k^{\frac{1}{2}}C_k^{\frac{3}{2}}} \sqrt{\frac{\big(-\sum_{j=i}^N \mu^*_{k,j}\big)^+}{\sum_{j=i}^N \lambda^*_{k,j}}},~~\forall i\in{\cal N}, k\in{\cal K} \\
 & l_{k,i}^{\rm off *} = \tau B\times \notag \\
 & \quad \log_2\Big(\max\Big[1,~ \frac{\sum_{j=i}^N\big(\nu^*_{j}-\mu^*_{k,j}\big)}{\sum_{j=i}^N\lambda^*_{k,j}}
\Big/\Big(\frac{\tau\sigma^2\ln2}{B\|\bm g_{k,i}\|^2}\Big)\Big]\Big),\notag \\
&\quad\quad\quad\quad\quad\quad\quad\quad\quad\quad \forall i\in{\cal N}\setminus\{N\}, k\in{\cal K}\\
 & \{\bm S_i^{*}\}\triangleq \argmin_{\{\bm S_i\succeq \bm 0\}} \sum_{i=1}^{N} \tau {\rm tr}(\bm S_i)\\
 &\quad\quad\quad\quad {\rm s.t.} ~ \sum_{j=1}^i\Big( \frac{\zeta_kC^3_k(l^{\rm loc*}_{k,j})^3}{\tau^2} + \frac{\tau \sigma^2(2^{\frac{l^{\rm off*}_{k,j}}{\tau B}}-1)}{\|{\bm g}_{k,j}\|^2} \Big) \notag \\
 & \quad\quad\quad\quad\quad\quad \leq \sum_{j=1}^{i} \tau\eta_k{\rm tr}(\bm S_j \bm h_{k,j} \bm h^H_{k,j}),~ \forall i\in{\cal N}, k\in{\cal K}, \notag
\end{align}
\end{subequations}
where $\lambda^*_{k,i}\geq 0$, $\mu^*_{k,j}\geq 0$, $\mu^*_{k,N}\in\mathbb{R}$, $\nu^*_j \geq 0$, and $\nu^*_N\in\mathbb{R}$ denote the optimal Lagrange multipliers associated with the constraints in (\ref{eq.prob1}b), (\ref{eq.prob1}c), (\ref{eq.prob1}d), (\ref{eq.prob1}e), and (\ref{eq.prob1}f), $j\in{\cal N}\setminus\{N\}$, $i\in{\cal N}$, $k\in{\cal K}$, respectively.
\end{theorem}
\begin{IEEEproof}
See Appendix A.
\end{IEEEproof}

Based on (\ref{eq.optimal-sol}a)--(\ref{eq.optimal-sol}b) in Theorem 1, we establish the following proposition regarding the optimal number of task input-bits allocated to the users' local computing and the AP's remote task execution.
\begin{proposition}\label{prop1}
For each user $k\in{\cal K}$ and the MEC server at the AP, the optimal number of executed task input-bits is, respectively, monotonically increasing over time; that is,
\begin{subequations}\label{eq.mono}
\begin{align}
&l^{\rm loc*}_{k,1} \leq \ldots \leq l^{\rm loc*}_{k,N}, ~~\forall k\in{\cal K} \\
&l^{\rm mec*}_{0,2} \leq \ldots \leq l^{\rm mec*}_{0,N}.
\end{align}
\end{subequations}
\end{proposition}
\begin{IEEEproof}
See Appendix B.
\end{IEEEproof}

To summarize, we present Algorithm 1 for optimally solving problem (P1) in the offline case in Table~I. Since the offline problem (P1) is a convex optimization problem and satisfies the Slater's condition, the strong duality holds between (P1) and its dual problem, and therefore, Algorithm~1 is guaranteed to converge to the global optimal solution to (P1)\cite{Boyd2004}. It takes no more than $2N^2K^2\log(RG/\epsilon)$ iterations in Algorithm 1 to achieve an $\epsilon$-optimal offline solution, where $R$ and $G$ denote the radius of the initial ellipsoid and the Lipschitz bound of the objective value of (P1), respectively\cite{Boyd14}.

\begin{table}[htp]\label{Alg1}
\begin{center}
\caption{Algorithm 1 for Optimally Solving Problem (P1) with TSI/CSI Known {\em a-priori}}
\begin{framed}
\begin{itemize}
\item[a)] {\bf Initialize} $\{\lambda_{k,i},\mu_{k,i},\nu_{i}\}$ with $\lambda_{k,i}\geq 0$, $\mu_{k,i}\geq 0$, $\nu_k\geq 0$, $\breve{\bm H}_i\triangleq \bm I_{N_t} - \sum_{k=1}^K \left(\sum_{j=i}^N\lambda_{k,j}\right)\eta_k\bm h_{k,i}\bm h_{k,i}^H \succeq \bm 0$, and $\sum_{j=1}^N \lambda_{k,j}>0$, $\forall k\in{\cal K}$, $i\in{\cal N}$.
\item[b)] {\bf Repeat:}
    \begin{itemize}
    \item[1)] Obtain $\{l^{\rm mec*}_{0,i}\}$, $\{l_{k,i}^{\rm loc*}\}$, and $\{l_{k,i}^{\rm off*}\}$ under given $\{\lambda_{k,i},\mu_{k,i},\nu_{i}\}$ according to Lemma~4, Lemma~5, and \eqref{eq.dual-sub3}, respectively (c.f. Appendix A);
    \item[2)] Update $\{\lambda_{k,i},\mu_{k,i},\nu_{i}\}$ based on the ellipsoid method\cite{Boyd2004}, where the subgradient of ${\cal G}(\{\lambda_{k,i},\mu_{k,i},\nu_{i})$ is $\Big[
        \frac{\zeta_1C_1^3(l_{1,1}^{\rm loc})^3}{\tau^2}+\frac{\tau\sigma^2}{\|\bm g_{1,1}\|^2}(2^{\frac{l^{\rm off*}_{1,1}}{\tau B}}-1),\ldots,\sum_{j=1}^N \frac{\zeta_KC_K^3(l_{k,j}^{\rm loc*})^3}{\tau^2}+\sum_{j=1}^N\frac{\tau\sigma^2}{\|\bm g_{K,j}\|^2}(2^{\frac{l^{\rm off*}_{K,j}}{\tau B}}-1), l^{\rm loc*}_{1,1}+ l^{\rm off*}_{1,1}-A_{1,1},\ldots,\sum_{j=1}^N l^{\rm loc*}_{K,j}+ \sum_{j=1}^N l^{\rm off*}_{K,j}-\sum_{j=1}^N A_{K,j},0,l^{\rm mec*}_{0,2}-\sum_{k=1}^K l^{\rm off*}_{K,1},\ldots,\sum_{j=1}^N l^{\rm mec*}_{0,j}-\sum_{j=1}^{N-1}\sum_{k=1}^K l^{\rm off*}_{k,j}\Big]^\dagger\in \mathbb{C}^{(2NK+N)\times 1}$ with respect to $\{\lambda_{k,i},\mu_{k,i},\nu_{i}\}$.
    \end{itemize}
\item[c)] {\bf Until} the dual variables $\{\lambda_{k,i},\mu_{k,i},\nu_{i}\}$ converge within the prescribed accuracy.
\item[d)] {\bf Set} $\{\lambda^{*}_{k,i},\mu^{*}_{k,i},\nu^{*}_{i}\} \leftarrow \{\lambda_{k,i},\mu_{k,i},\nu_{i}\}$.
\item[e)] {\bf Output}:
Obtain $\{ l_{0,i}^{\rm mec*}\}$, $\{ l^{\rm loc*}_{k,i}\}$, and $\{l^{\rm off*}_{k,i}\}$ under the dual optimal $\{\lambda^{*}_{k,i},\mu^{*}_{k,i},\nu^{*}_{i}\}$ according to (\ref{eq.optimal-sol}a), (\ref{eq.optimal-sol}b), and (\ref{eq.optimal-sol}c), respectively, and compute $\{\bm S_i^{*}\}$ by solving the SDP in (\ref{eq.optimal-sol}d) in Theorem 1.
\end{itemize}
\end{framed}
\end{center}\label{algorithm:new0}
\end{table}

\begin{remark}
Proposition \ref{prop1} reveals the {\em monotonically increasing} feature of the optimal number of executed task input-bits for both the users and the AP. This feature can be intuitively understood as follows. Each user $k$'s computation energy consumption function in \eqref{eq.Eki} is identical over different slots and it is also convex with respect to the number of task input-bits, and so is the AP's computation energy consumption. Therefore, in order to reduce the energy consumption, both the users and AP should distribute their computation tasks as evenly as possible over time. Due to the task causality constraints in \eqref{eq.Si-user} and \eqref{eq.MECi}, more computation tasks will be accumulated at the users and AP over time, and therefore the computation load at the users/AP should monotonically increases over time. This monotonic feature is also reminiscent of the monotonically increasing power allocation in wireless energy harvesting based communications due to the energy causality constraints (see, e.g., \cite{Feng19,Li15,Rui12}).
\end{remark}

Based on (\ref{eq.optimal-sol}c) in Theorem 1, the optimal number of offloaded task input-bits $l_{k,i}^{\rm off*}$ for user $k\in{\cal K}$ at slot $i\in{\cal N}$ should be adapted according to the task load at the AP, as well as the CSI for WPT and offloading. In a special case where the CSI for both WPT and offloading remains unchanged over slots for user $k\in{\cal K}$, if $\nu^*_i \leq \mu^*_{k,i}$, $\forall i\in{\cal N}$, then the optimal number of task input-bits allocated for offloading is monotonically increasing over time, i.e., $l^{\rm off*}_{k,i}\leq \ldots \leq l^{\rm off*}_{k,N-1}$. This result indicates that user $k$ should offload more tasks at the next slot $(i+1)$ if the penalty (corresponding to the weighted cumulative difference $\sum_{j=i}^N(\nu^*_j-\mu_{k,j})/(\sum_{j=i}^N\lambda^*_{k,j})$) caused by the AP's energy consumption in violating the task causality constraint is smaller than that by user $k$ at slot $i$, $\forall i\in\{1,\ldots,N-2\}$. In addition, each user $k\in{\cal K}$ is supposed to adapt its offloaded tasks to the offloading channel power gain $\|\bm g_{k,i}\|^2$ at slot $i\in{\cal N}\setminus\{N\}$.

\begin{example}[Single User Case]
We now consider a single user case with $K=1$. To start with, we denote $i^*\in{\cal N}$ as the index of the {\em causality-dominating} WPT slot up to the present slot $i$, which corresponds to the largest WPT channel power gain and is defined as
\begin{align}\label{eq.i-star}
i^* \triangleq \argmax_{1\leq j \leq i} \|\bm h_{1,j}\|^2.
\end{align}
Based on \eqref{eq.i-star} and Theorem \ref{thorem1}, the optimal solution to problem (P1) with $K=1$ is obtained as
\begin{subequations} \label{eq.sol-single-user}
\begin{align}
& l_{0,i}^{\rm mec*} = \tau\sqrt{\frac{\big(-\sum_{j=i}^N{\nu}^*_j\big)^{+}}{3\zeta_0C_0^3}},~~\forall i\in{\cal N}\setminus\{1\} \\
& l_{1,i}^{\rm loc*} = \tau\|\bm h_{1,i^*}\|\sqrt{\frac{\big(-\sum_{j=i}^N{\mu}^*_{1,j}\big)^+}{3\eta_1\zeta_1C_1^3}},~~\forall i\in{\cal N} \\
& l_{1,i}^{\rm off*} = \tau B \times \notag\\
&~ \log_2\Big( \max\Big[1,\Big(\sum_{j=i}^N{\nu}^*_j-\sum_{j=i}^N{\mu}^*_{1,j}\Big)\Big/{\frac{\tau \sigma^2\ln2}{\eta_1 B \|\bm h_{1,i^*}\|\|\bm g_{1,i}\|^2}} \Big]\Big),\notag\\
&\quad\quad\quad\quad\quad\quad\quad\quad\quad\quad \forall i\in{\cal N}\setminus\{N\} \\
&\bm S^*_i = \begin{cases}
\frac{p^*_{i^*}\bm h_{1,i^*} \bm h^H_{1,i^*}} { \|\bm h_{1,i^*}\|^2}, & {\rm if}~ i=i^*, \forall i\in{\cal N}\\
\bm 0,& {\rm otherwise},
\end{cases}
\end{align}
\end{subequations}
where
\begin{align}\label{eq.sum}
&p^*_{i^*} =  \frac{1}{N_{i^*} \tau \eta_1 \|\bm h_{1,i^*}\|^2}\times \notag \\
&~~~~ \sum_{j=i^*}^{(i+N_{i^*}-1)^*}\Big(\frac{\zeta_1C_1^3}{\tau^2}(l_{1,j}^{\rm loc*})^3 + \frac{\tau \sigma^2}{\|\bm g_{1,j}\|^2} (2^{\frac{l_{1,j}^{\rm off*}}{\tau B}}-1) \Big)
\end{align}
with $N_{i^*}$ (satisfying $(i^*+N_{i^*}-1)^*=i^*$ and $(i^*+N_{i^*})^*=i^*+1$) being defined as the number of slots between the $i^*$-th and $(i^*+1)$-th causality-dominating WPT slots.

\begin{remark}
Based on \eqref{eq.sol-single-user}, some interpretations on the optimal offline wireless powered MEC design in the single-user case are stated in the following.
\begin{itemize}
\item First, in line with (\ref{eq.sol-single-user}a), since the computation level $\bar{\nu}_i\triangleq-\sum_{j=i}^N{\nu}^*_j\geq 0$ is monotonically increasing over time, the number of task input-bits allocated for the AP's remote execution is monotonically increasing over time.
\item Second, based on the monotonic structure of the causality-dominating WPT channel gains (i.e., $\| \bm h_{1,1^*}\| \leq \|\bm h_{1,2^*}\|\leq \ldots \leq \|\bm h_{1,N^*}\|$) and the monotonically increasing computation level for local computing (i.e., $\bar{\mu}_{1,1}\leq \ldots\leq \bar{\mu}_{1,N}$ with $\bar{\mu}_{1,i}\triangleq-\sum_{j=i}^N{\mu}^*_{1,j}$, $\forall i\in{\cal N}$), it follows from (\ref{eq.sol-single-user}b) that $l_{1,1}^{\rm loc*}\leq\ldots\leq l_{1,N}^{\rm loc*}$. This implies that the user can harvest a larger amount of energy at the future WPT causality-dominating slots than at the earlier ones. Therefore, the user prefers to execute more tasks at the future causality-dominating WPT slot intervals for energy saving.
\item Third, as shown in (\ref{eq.sol-single-user}c), the optimal number of task input-bits for the user's task offloading depends on both the user's local computing and AP's remote task execution, as well as the channel power gains for offloading and WPT at the associated causality-dominating slots. With the larger value of the difference $(\bar{\mu}_{1,i}-\bar{\nu}_i)$, as well as the larger channel power gains for offloading and WPT, the user prefers to offload more tasks to the AP.
\item Finally, (\ref{eq.sol-single-user}d) indicates that the MRC based energy beamforming is optimal for WPT, and the AP only needs to transfer wireless energy at those causality-dominating WPT slots. As shown in \eqref{eq.sum}, the optimal transmit-power amount $p^*_{i^*}$ of the AP at slot $i^*$ is exactly equal to that demanded by the user's computation before the next causality-dominating WPT slot.
\end{itemize}
\end{remark}
\end{example}

\section{Proposed Online Design for Solving (P1)}\label{sec:optimal}
In the previous section, we have obtained the offline solution to problem (P1) under the assumption that the TSI/CSI ($\{A_{k,i},\bm h_{k,i},\bm g_{k,i}\}_{i=1}^N$) are perfectly known {\em a-priori}. Inspired by the optimal offline solution in Theorem~1, in this section we solve problem (P1) in real time with the causally known TSI/CSI. With this regard, at each slot $i\in{\cal N}$, the perfect TSI/CSI, $\{A_{k,j},\bm h_{k,j},\bm g_{k,j}\}_{j=1}^i$, and the predicted ones, $\{\hat{A}_{k,j},\hat{\bm h}_{k,j},\hat{\bm g}_{k,j}\}_{j=i+1}^N$, are known at the AP/users. The main thrust of our proposed online scheme is to combine the optimal offline solution for problem (P1) with a sliding-window based sequential optimization \cite{Rahbar15,Cama13}.

To start with, we define an integer parameter $M\in\{1,\ldots,N\}$ as the size of the chosen sliding-window. At each slot $i\in{\cal N}$, we view the online optimization problem as a finite-horizon energy minimization problem over a window of $M$ slots, with the available task arrivals over this window as $\{A_{k,i},\hat{A}_{k,i+1},\ldots,\hat{A}_{k,i+M-1}\}$, the available CSI for WPT as $\{\bm h_{k,i},\hat{\bm h}_{k,i+1},\ldots,\hat{\bm h}_{k,i+M-1}\}$, and the available CSI for offloading as $\{\bm g_{k,i},\hat{\bm g}_{k,i+1},\ldots,\hat{\bm g}_{k,i+M-1}\}$. Note that the window of size $M$ should not exceed the $N$-slot horizon, i.e., it must hold that $i+M-1\leq N$. If $i+M-1>N$, then we decrease the window size as $M=N-i+1$, such that the corresponding window consists of slots $\{i,i+1,\ldots,N\}$. For the online optimization problem at slot $i$, we denote the design variables over the present window of size $M$ as $\{{\bm S}_j^{(i)},l_{0,j}^{{\rm mec}(i)},l_{k,j}^{{\rm loc}(i)},l_{k,j}^{{\rm off}(i)}\}_{j=1}^M$. Then, we formulate problem (P1-SW($i$)) at slot $i$ (c.f.~\eqref{eq.prob_OPi}) similarly as problem (P1), by replacing $N$ in (P1) by $M$, $\{A_{k,j}\}_{j=1}^N$ by $\{A_{k,i},\hat{A}_{k,i+1},\ldots,\hat{A}_{k,i+M-1}\}$, $\{\bm h_{k,j}\}_{j=1}^N$ by $\{\bm h_{k,i},\hat{\bm h}_{k,i+1},\ldots,\hat{\bm h}_{k,i+M-1}\}$, $\{\bm g_{k,j}\}_{j=1}^N$ by $\{\bm g_{k,i},\hat{\bm g}_{k,i+1},\ldots,\hat{\bm g}_{k,i+M-1}\}$, and finally the design variables $\{{\bm S}_j,l_{0,j}^{{\rm mec}},l_{k,j}^{{\rm loc}},l_{k,j}^{{\rm off}}\}_{j=1}^N$ by $\{{\bm S}_j^{(i)},l_{0,j}^{{\rm mec}(i)},l_{k,j}^{{\rm loc}(i)},l_{k,j}^{{\rm off}(i)}\}_{j=1}^M$. Furthermore, we denote $\{\tilde{\bm S}_j,\tilde{l}_{0,j}^{{\rm mec}},\tilde{l}_{k,j}^{{\rm loc}},\tilde{l}_{k,j}^{{\rm off}}\}_{j=1}^N$ as the solution obtained by the proposed sliding-window online scheme, and define
\begin{align}
R^{\rm mec}_{0,i} & \triangleq \sum_{j=1}^{i-1} \sum_{k=1}^K \tilde{l}_{k,j}^{{\rm off}} - \sum_{j=1}^{i-1} \tilde{l}_{0,j}^{\rm mec},~~\forall i\in{\cal N}\\
R_{k,i} & \triangleq \sum_{j=1}^{i-1}A_{k,j} - \sum_{j=1}^{i-1} (\tilde{l}_{k,j}^{{\rm loc}} + \tilde{l}_{k,j}^{\rm off}), ~~\forall i\in{\cal N},k\in{\cal K}
\end{align}
as the residual number of task input-bits at the AP and user $k\in{\cal K}$ at the beginning of slot $i$, respectively. Without loss of generality, it is assumed that $R^{\rm mec}_{0,1}=0$ and $R_{k,1}=0$, $\forall k\in{\cal K}$. Accordingly, the online problem~(P1-SW($i$)) with the chosen window size $M$ is formulated as
\begin{subequations}\label{eq.prob_OPi}
\begin{align}
&({\tt P1-SW}(i)):\notag \\
&\min_{\{\bm S_j^{(i)},l^{{\rm mec}(i)}_{0,j},l^{{\rm loc}(i)}_{k,j},l^{{\rm off}(i)}_{k,j}\}} \sum_{j=1}^{M} \Big( \tau{\rm tr}(\bm S^{(i)}_j) + \frac{\zeta_0 C_0^3 (l^{{\rm mec}(i)}_{0,j})^3}{\tau^2} \Big)\\
&~{\rm s.t.}~  \frac{\tau\sigma^2(2^{\frac{l^{{\rm off}(i)}_{k,m}}{\tau B}}-1)}{\|{\bm g}_{k,i}\|^2} + \sum_{m=1}^j \frac{\zeta_k C_k^3 (l^{{\rm loc}(i)}_{k,m})^3}{\tau^2} \notag \\
&\quad + \sum_{m=2}^j \frac{\tau\sigma^2(2^{\frac{l^{{\rm off}(i)}_{k,m}}{\tau B}}-1)}{\|\hat{\bm g}_{k,i+m-1}\|^2} \leq  \tau \eta_k {\rm tr}(\bm S^{(i)}_1\bm h_{k,i}\bm h_{k,i}^H) + \notag \\
&\quad\quad \sum_{m=2}^j \tau \eta_k {\rm tr}(\bm S^{(i)}_m\hat{\bm h}_{k,i+m-1}\hat{\bm h}^H_{k,i+m-1}),~~\forall j\in {\cal M},k \in{\cal K}\\
&\quad
R_{k,i} + A_{k,i} - l^{{\rm loc}(i)}_{k,1} - l^{{\rm off}(i)}_{k,1} \geq 0 \\
&\quad R_{k,i} + A_{k,i} + \sum_{m=2}^j \hat{A}_{k,i+m-1} - \sum_{m=1}^j (l^{{\rm loc}(i)}_{k,m} + l^{{\rm off}(i)}_{k,m}) \geq 0,\notag \\
&\quad\quad\quad\quad\quad\quad\quad\quad\quad\quad\quad \forall j\in {\cal M}\setminus\{1,M\},~ k \in{\cal K}\\
&\quad
R_{k,i} + A_{k,i} + \sum_{m=2}^{M} \hat{A}_{k,i+m-1} - \sum_{m=1}^M (l^{{\rm loc}(i)}_{k,m} + l^{{\rm off}(i)}_{k,m}) = 0,\notag \\
& \quad\quad\quad\quad\quad\quad\quad\quad\quad\quad\quad\quad\quad\quad\quad \forall k\in {\cal K}\\
&\quad
R^{\rm mec}_{0,i} - l^{{\rm mec}(i)}_{0,1} \geq 0 \\
&\quad
R^{\rm mec}_{0,i} + \sum_{m=1}^{j-1} \sum_{k=1}^K l^{{\rm off}(i)}_{k,m} - \sum_{m=1}^j l^{{\rm mec}(i)}_{0,m} \geq 0, \notag \\
& \quad\quad\quad\quad\quad\quad\quad\quad\quad\quad\quad\quad\quad\quad \forall j\in {\cal M}\setminus\{1,M\}\\
&\quad
R^{\rm mec}_{0,i} + \sum_{m=1}^{M-1} \sum_{k=1}^K l^{{\rm off}(i)}_{k,m} - \sum_{m=1}^M l^{{\rm mec}(i)}_{0,m} = 0\\
&\quad \bm S_j^{(i)}\succeq \bm 0, l^{{\rm mec}(i)}_{0,j}\geq 0, l^{{\rm loc}(i)}_{k,j}\geq 0, l^{{\rm off}(i)}_{k,j}\geq 0,\notag\\
&\quad\quad\quad\quad\quad\quad\quad\quad\quad\quad\quad\quad\quad\quad\quad \forall j\in{\cal M}, k\in{\cal K}.
\end{align}
\end{subequations}
Problem (P1-SW($i$)) can be efficiently solved by Algorithm 1 by a change of variables/parameters as specified above. Note that it does not necessarily hold that $l_{0,1}^{{\rm mec}(i)}=0$ and $l_{k,M}^{{\rm off}(i)}=0$ due to the nonnegative residual number of task input-bits $R_{0,i}^{\rm mec}\geq 0$ for (P1-SW($i$)), $\forall i\in{\cal N}\setminus\{1,N\}$.

By denoting the optimal solution to problem (P1-SW($i$)) as $\{{\bm S}^{(i)*}_j,l_{0,j}^{{\rm mec}(i)*},l_{k,j}^{{\rm loc}(i)*},l_{k,j}^{{\rm off}(i)*}\}_{j=1}^M$, $\forall i\in{\cal N}$, the solution of the proposed online sliding-window based scheme is obtained as
\begin{align} \label{sol_online}
&\tilde{\bm S}_i = \bm S_1^{(i)*}, ~\tilde{l}_{0,i}^{{\rm mec}}=l_{0,1}^{{\rm mec}(i)*},~\tilde{l}_{k,i}^{{\rm loc}} =l_{k,1}^{{\rm loc}(i)*}, ~ \tilde{l}_{k,i}^{{\rm off}} = l_{k,1}^{{\rm off}(i)*},\notag \\
&~~~~~~~~~~~~~~~~~~~~~~~~~~~~~~~~~~~~~~~~~~~~~\forall i\in{\cal N},k\in{\cal K}.
\end{align}

In summary, we present the proposed online sliding-window based scheme as Algorithm~2 in Table~II. For problem (P1) in the online case with a chosen sliding-window size $M$, it involves sequentially solving $N$ convex optimization problems, and therefore, the convergence of each problem (P1-SW($i$)), $\forall i\in{\cal N}$, is guaranteed\cite{Boyd2004}. In line with the analysis for Algorithm~1, it takes no more than $2M^2K^2\log(RG/\epsilon)$ iterations to obtain an $\epsilon$-optimal solution to (P1-SW($i$)), and thus the total number of iterations is no more than $2NM^2K^2\log(RG/\epsilon)$ in Algorithm~2~\cite{Boyd14}.

\begin{table}[htp]\label{Alg2}
\begin{center}
\caption{Algorithm 2 of the Proposed Online Sliding-Window based Scheme for Problem (P1)}
\begin{framed}
\begin{itemize}
\item[a)] {\bf Initialize} $i\gets 1$, $R_{k,1}=0$, $\forall k\in{\cal K}$, $R^{\rm mec}_{0,1}=0$, and $1\leq M \leq N$.
\item[b)] {\bf Repeat:}
    \begin{itemize}
    \item[1)]  For slot $i$, solve (P1-SW($i$)) by Algorithm~1, and obtain its solution as $\{{\bm S}^{(i)*}_j,l_{0,j}^{{\rm mec}(i)*},l_{k,j}^{{\rm loc}(i)*},l_{k,j}^{{\rm off}(i)*}\}_{j=1}^M$;
    \item[2)]  Set $\tilde{\bm S}_i = \bm S_1^{(i)*}$, $\tilde{l}_{0,i}^{{\rm mec}}=l_{0,1}^{{\rm mec}(i)*}$, $\tilde{l}_{k,i}^{{\rm loc}} =l_{k,1}^{{\rm loc}(i)*}$, and $\tilde{l}_{k,i}^{{\rm off}} = l_{k,1}^{{\rm off}(i)*}$, $\forall k\in{\cal K}$ (c.f.~\eqref{sol_online});
    \item[3)] Set $i\gets i+1$;

    \item[4)] Set $R_{k,i} = \sum_{j=1}^{i-1}A_{k,j} - \sum_{j=1}^{i-1}(\tilde{l}_{k,j}^{{\rm loc}} + \tilde{l}_{k,j}^{\rm off})$ and $ R^{\rm mec}_{0,i}=\sum_{j=1}^{i-1} \sum_{k=1}^K \tilde{l}_{k,j}^{{\rm off}} - \sum_{j=1}^{i-1} \tilde{l}_{0,j}^{\rm mec}$.
      \end{itemize}
\item[c)] {\bf Until} $i=N+1$.
\item[e)] {\bf Output}: The online sliding-window based solution is $\{\tilde{\bm S}_i,\tilde{l}_{0,i}^{{\rm mec}},\tilde{l}_{k,i}^{{\rm loc}},\tilde{l}_{k,i}^{{\rm off}}\}_{i=1}^N$.
\end{itemize}
\end{framed}
\end{center}
\vspace{-0.6cm}
\end{table}

\begin{remark}
In the proposed online sliding-window based scheme, the sliding-window size $M$ plays a key role in balancing the TSI/CSI predication error and the system performance\cite{Rahbar15,Cama13}. Specifically, the benefit of a large sliding-window size $M$ is reflected in the cases with small prediction errors by fully exploiting the long-term TSI/CSI predication. On the other hand, since the predication of TSI/CSI becomes less useful as the sliding-window size $M$ decreases, preferences to smaller size $M$ are reflected in the cases with large TSI/CSI predication errors. In Section VI-B, we also numerically corroborated the effects of the sliding-window size versus the TSI/CSI prediction error. Compared with the online optimal DP method\cite{DP}, the proposed online sliding-window based scheme does not necessarily achieve the optimal online performance but maintains a low computational complexity in general. It is worthy of further investigation of quantified relation between the sliding-window size $M$ and the TSI/CSI errors in future work.
\end{remark}

\subsection{Special Case with $M=1$: Myopic Design}
In this subsection, we consider a special case by setting the window size as $M=1$, which is also known as {\em myopic} design \cite{Rahbar15}. In this myopic design scheme, each user $k\in{\cal K}$ needs to complete the number of $A_{k,i}$ task input-bits at slot $i\in{\cal N}$ (via local computing and/or offloading), and the AP needs to execute the number of the $\sum_{k=1}^K l^{\rm off}_{k,i-1}$ task input-bits which are offloaded by the $K$ users during previous slot $(i-1)$. Denote $\{\hat{\bm S}_i,\hat{l}_{0,i}^{\rm mec}, \hat{l}_{k,i}^{\rm loc}, \hat{l}_{k,i}^{\rm off}\}$ as the myopic design based solution to problem (P1). Herein, we establish the following proposition.

\begin{proposition} \label{Proposition-myopic}
The myopic design based solution $\{\hat{\bm S}_i,\hat{l}_{0,i}^{\rm mec}, \hat{l}_{k,i}^{\rm loc}, \hat{l}_{k,i}^{\rm off}\}$ to (P1) is given as
\begin{subequations}\label{Proposition-myopic-sol}
\begin{align}
& \hat{l}_{k,i}^{\rm loc} = A_{k,i} - \hat{l}_{k,i}^{\rm off},~~\forall i\in{\cal N}, k\in{\cal K} \\
& \hat{l}_{0,i}^{\rm mec} = \sum_{k=1}^K \hat{l}_{k,i}^{\rm off},~~ \forall i\in{\cal N}\setminus\{1\}\\
& \hat{\bm S}_i = \argmin_{\bm S_i\succeq \bm 0}~~{\rm tr}(\bm S_i) \notag \\
& \quad\quad\quad {\rm s.t.}~~ \frac{\zeta_kC_k^3(\hat{l}_{k,i}^{\rm loc})^3}{\tau^2} + \frac{\tau \sigma^2}{\|\bm g_{k,i}\|^2}(2^{\frac{\hat{l}_{k,i}^{\rm off}}{\tau B}}-1) \notag \notag \\
&\quad\quad\quad\quad\quad\quad \leq \tau \eta_k {\rm tr}(\bm S_i \bm h_{k,i} \bm h_{k,i}^H),~~\forall k\in{\cal K}
\end{align}
\end{subequations}
for $i\in{\cal N}$, where $\hat{l}_{k,i}^{\rm off}$ satisfies $\frac{3\zeta_kC_k^3(A_{k,i}-\hat{l}_{k,i}^{\rm off})^2}{\tau^2} = \frac{\tau \sigma^2 \ln2}{\|\bm g_{k,i}\|^2} 2^{\frac{\hat{l}_{k,i}^{\rm off}}{\tau B}}$, $\forall i\in{\cal N}\setminus\{N\}$, and $\hat{l}_{k,N}^{\rm off}=0$, $\forall k\in{\cal K}$.
\end{proposition}
\begin{IEEEproof}
See Appendix C.
\end{IEEEproof}

\section{Numerical Results}
In this section, we provide numerical results to evaluate the performance of the proposed joint-WPT-MEC designs for a wireless powered multiuser MEC system. Denoting $d_k$ as the distance between user $k$ and the AP, we consider the following distance-dependent Rician fading channel models\cite{Goldsmith}:
\begin{subequations}
\begin{align} \label{eq.rician_channel}
& \bm h_{k,i} = \sqrt{\frac{{\cal X}_R\Omega_0d_k^{-\alpha}}{1+{\cal X}_R}}\bm h_0 + \sqrt{\frac{\Omega_0d_k^{-\alpha}}{1+{\cal X}_R}} \bm h \\
& \bm g_{k,i} = \sqrt{\frac{{\cal X}_R\Omega_0d_k^{-\alpha}}{1+{\cal X}_R}}\bm g_0 + \sqrt{\frac{\Omega_0d_k^{-\alpha}}{1+{\cal X}_R}} \bm g
\end{align}
\end{subequations}
for all $i\in{\cal N}$ and $k\in{\cal K}$, where ${\cal X}_R$ denotes the Rician factor, $\Omega_0$ dB corresponds to the path loss at a reference distance of one meter (m), $\alpha$ is the pathloss exponent, the line-of-sight (LoS) components $\bm h_0$ and $\bm g_0$ have all elements equal to one, and both $\bm h\in\mathbb{C}^{N_t\times 1}$ and $\bm g\in\mathbb{C}^{N_t\times 1}$ are randomly generated CSCG vectors $\sim{\cal CN}(\bm 0,\bm I_{N_t})$. In the simulations, unless stated otherwise, the system parameters are listed in Table III.

\begin{table*}
\centering
\caption{Simulation Parameters}
\begin{tabular}{|l|l|l|l|}
\hline
$\eta_k=0.3,\forall k\in{\cal K}$ & $C_0=C_k=10^3$ CPU cycles/bit, $\forall k\in{\cal K}$ &  $N_t=4$ &  $\alpha=3$ \\
$\zeta_0=10^{-29}$ & $A_{k,i}\sim {\cal U}[5\times 10^5,~10^6]$ bits, $\forall k\in{\cal K}, i\in{\cal N}$ & $\sigma^2=10^{-9}$ Watt & ${\cal X}_R=3$ \\
$B=2$ MHz & $\sigma_A=\sigma_h=\sigma_g=0.2$ & $\zeta_k=10^{-28}$, $\forall k\in{\cal K}$ & $\Omega_0=-32$ dB\\
\hline
\end{tabular}
\end{table*}

\subsection{Offline Designs}

\begin{figure}
\begin{minipage}[c]{0.5\textwidth}
  \centering
  \includegraphics[width = 3.3in]{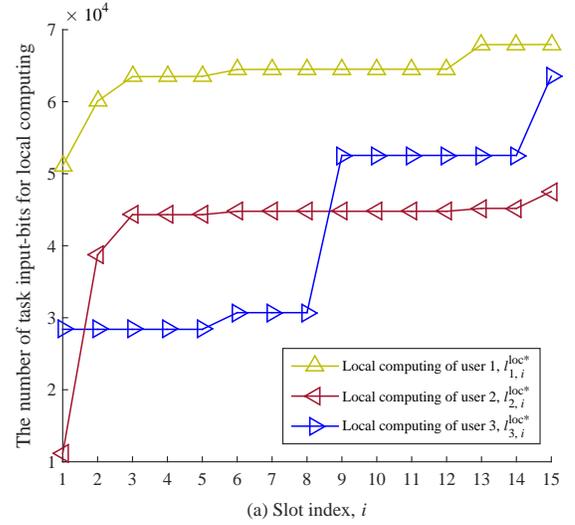}
\end{minipage}
\begin{minipage}[c]{0.5\textwidth}
  \centering
  \includegraphics[width = 3.3in]{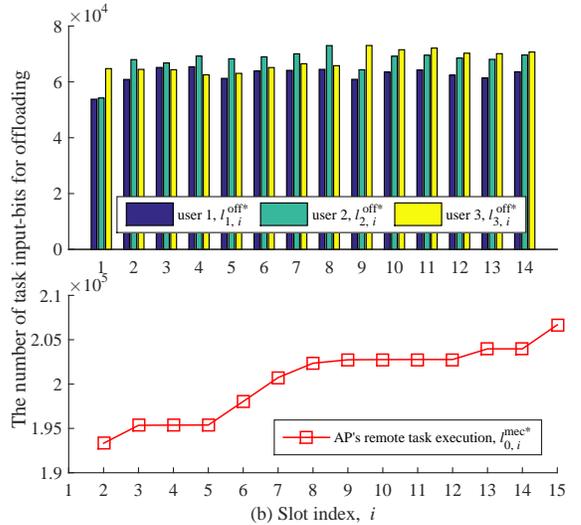}
  \end{minipage}
\caption{The number of task-input bits allocated for the users' local computing and offloading, as well as the AP's remote task execution at different slots.} \label{fig.Offline}
\end{figure}

First, we consider the optimal offline solution to problem (P1) with the TSI/CSI perfectly known {\em a-priori}, where the user number is set as $K=3$, the time horizon consists of $N=15$ slots each with duration $\tau_i=0.02$ second (sec), $\forall i\in{\cal N}$, and the distance between user $k$ and the AP is set as $d_k=3$ m, $\forall k\in{\cal K}$. The number $A_{k,i}$ of task input-bits for user $k\in{\cal K}$ at slot $i\in{\cal N}$ is set as $A_{k,i}\sim {\cal U}[5\times 10^5,10^6]$. Under the setup, Fig.~\ref{fig.Offline} shows the optimal $\{l^{\rm mec*}_{0,i},l_{k,i}^{\rm loc*},l_{k,i}^{\rm off*}\}$ to problem (P1) with one channel realization. It is observed in Fig.~\ref{fig.Offline}(a) that the optimal number of task input-bits $l^{\rm loc*}_{k,i}$ allocated to user $k$'s local computing is monotonically increasing over slots, which corroborates~(\ref{eq.mono}a) in Proposition~\ref{prop1}. Interestingly, we observe that from the 9-th slot on, user 3 starts to execute a significantly increased number of task input-bits for local-computing, which is because the number $A_{3,9}$ of task input-bits is significantly larger than those arriving at previous slots. In Fig.~\ref{fig.Offline}(b), the optimal number of task input-bits $l^{\rm off*}_{k,i}$ allocated for offloading from user $k$ is observed to be proportional to its channel power gain for task offloading $\|\bm g_{k,i}\|^2$. Specifically, a larger $\|\bm g_{k,i}\|^2$ value leads to a larger $l_{k,i}^{\rm off*}$ value and vice versa, which is consistent with the semi-closed form of $l_{k,i}^{\rm off*}$ in (\ref{eq.optimal-sol}c) in Theorem~1. It is also expected in Fig.~\ref{fig.Offline}(b) that the optimal number of task-input bits $l^{\rm mec*}_{0,i}$ allocated for remote execution at the AP is monotonically increasing over slots, which is expected and consistent with (\ref{eq.mono}b) in Proposition~\ref{prop1}.

For comparison, we consider the following three benchmark schemes for offline wireless powered MEC designs of problem (P1).
\begin{itemize}
\item {\em Local computing only:} Each user $k\in{\cal K}$ needs to accomplish its computation tasks by only local computing. This scheme corresponds to solving problem (P1) by setting $l^{\rm mec}_{0,i}=0$ and $l^{\rm off}_{k,i}=0$, $\forall i\in{\cal N}$, $k\in{\cal K}$.
\item {\em Full offloading:} Each user $k\in{\cal K}$ needs to accomplish its computation tasks by fully offloading them to the AP. This scheme corresponds to solving problem (P1) by setting $l^{\rm loc}_{k,i}=0$, $\forall i\in{\cal N}$, $k\in{\cal K}$.
\item {\em Myopic design:} At each slot $i\in{\cal N}$, both user $k\in{\cal K}$ and the AP need to accomplish their task input-bits of $A_{k,i}$ and $\sum_{k=1}^K l^{\rm off}_{k,i-1}$, respectively, i.e., we have $A_{k,i}-l^{\rm loc}_{k,i}-l^{\rm off}_{k,i}=0$ and $l^{\rm mec}_{0,i}-\sum_{k=1}^K l^{\rm off}_{k,i-1}=0$, $\forall i\in{\cal N}$, $k\in{\cal K}$. In this scheme, the total system energy minimization can be implemented independently over each individual time slot, and the optimal solution is obtained as in Proposition~2.
\end{itemize}

\begin{figure}
  \centering
  \includegraphics[width = 3.3in]{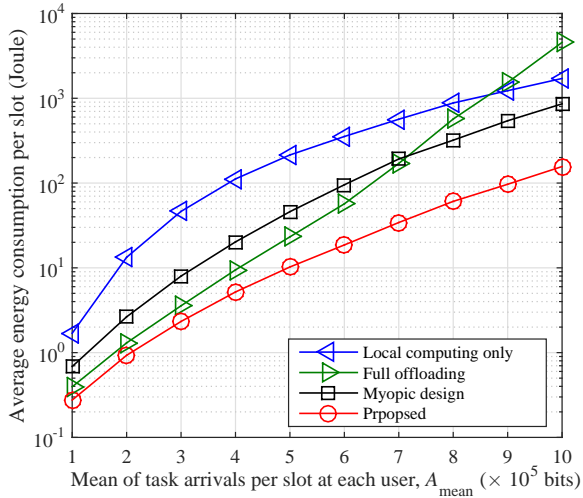}
 \caption{The average energy consumption per slot at the AP versus the mean number of task input-bits $A_{\rm mean}$.} \label{fig.vsAmax}
\end{figure}

Fig.~\ref{fig.vsAmax} shows the average energy consumption at the AP per slot versus the mean of the task input-bits $A_{\rm mean}$, where $\tau=0.02$ sec, $N=20$, $K=6$, and $d_k=4$ m, $\forall k\in{\cal K}$. The amount of dynamic task arrivals is set to follow a uniform distribution with $A_{k,i}\sim {\cal U}[0,2A_{\rm mean}]$ Mbits, $\forall k\in{\cal K}$, $i\in{\cal N}$, and the numerical results are obtained by averaging over $1000$ randomized channel realizations and randomized task realizations. It is observed that the proposed design achieves the least average energy consumption among all the four schemes, and its gain over the benchmark schemes becomes more significant as $A_{\rm mean}$ increases. With small value of $A_{\rm mean}$ (e.g., $A_{\rm mean}\leq 7$ Mbits), the benchmark full-offloading scheme is observed to outperform the myopic-design scheme, which indicates the advantage of considering task arrivals from the perspective of the whole horizon. It is also observed that the full-offloading scheme performs inferior to myopic-design scheme and local-computing-only scheme with large $A_{\rm mean}$ values. This is due to the fact that the energy consumption for offloading increases more significantly (exponentially) than that for local computing (cubically).

\begin{figure}
  \centering
  \includegraphics[width = 3.3in]{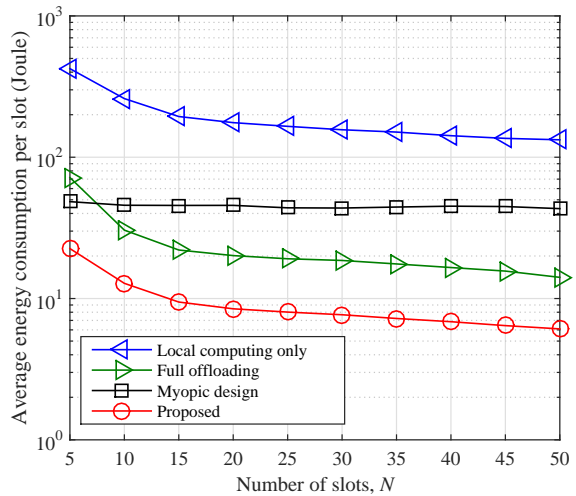}
 \caption{The average energy consumption per slot at the AP versus the number $N$ of slots.} \label{fig.vsN}
\end{figure}

Fig.~\ref{fig.vsN} shows the average energy consumption per slot at the AP versus the number $N$ of slots within the horizon, where $\tau=0.02$ sec, $K=4$, and $d_k= 4$ m, $\forall k\in{\cal K}$. The amount of dynamic task arrivals is set to follow the uniform distribution with $A_{k,i}\sim {\cal U}[0,5]$ Mbits, $\forall k\in{\cal K}$, $i\in{\cal N}$, and the numerical results are obtained by averaging over $1000$ randomized channel realizations and randomized task realizations. It is observed that the proposed offline design achieves the lowest average energy consumption among all the four schemes, which validates the importance of joint local-computing and offloading design over multiple time slots in terms of energy saving. The myopic design is observed to achieve almost unchanged average energy consumption per slot, which is due to the fact that the users and the AP are shortsighted by only considering tasks related to the present or last slot. By contrast, all the other design schemes are observed to admit decreasing per-slot energy consumption. This is intuitively expected, since as $N$ increases, the users and the AP have more degrees of freedom in scheduling the task allocation for energy saving. In addition, the local-computing-only scheme is observed to perform inferior to the other two benchmark schemes in this setup.

\subsection{Online Designs}
Next, we evaluate the performance of the proposed online sliding-window based scheme for problem (P1). In the following simulations, we consider the relative predication error of the TSI at slot $i$ as $\delta^\prime_{A_{k,i}}={\delta_{A_{k,i}}}/{A_{k,i}}$, where $\delta^\prime_{A_{k,i}}\sim {\cal N}(0,\sigma_A^2)$, $\forall i\in{\cal N}$, $\forall k\in{\cal K}$. Likewise, we define the relative prediction errors of CSI for WPT and for task offloading between user $k\in{\cal K}$ and the AP at slot $i\in{\cal N}$ as $\bm \delta^\prime_{h_{k,i}} \triangleq \bm \delta_{h_{k,i}}\Big/\Big( \sqrt{\frac{\Omega_0 d_k^{-\alpha}}{1+{\cal X}_R}}\bm h \Big)$ and $\bm \delta^\prime_{g_{k,i}} \triangleq {\bm \delta}_{g_{k,i}}\Big/\Big(\sqrt{\frac{\Omega_0 d_k^{-\alpha}}{1+{\cal X}_R}}\bm g\Big)$, respectively, where ${\bm \delta}^\prime_{h_{k,i}}\sim{\cal CN}(\bm 0,\sigma^2_h\bm I_{N_t})$ and ${\bm \delta}^\prime_{g_{k,i}}\sim{\cal CN}(\bm 0,\sigma^2_g\bm I_{N_t})$, $\forall i\in{\cal N}$, $\forall k\in{\cal K}$.

\begin{figure}
  \centering
  \centering
  \includegraphics[width = 3.3in]{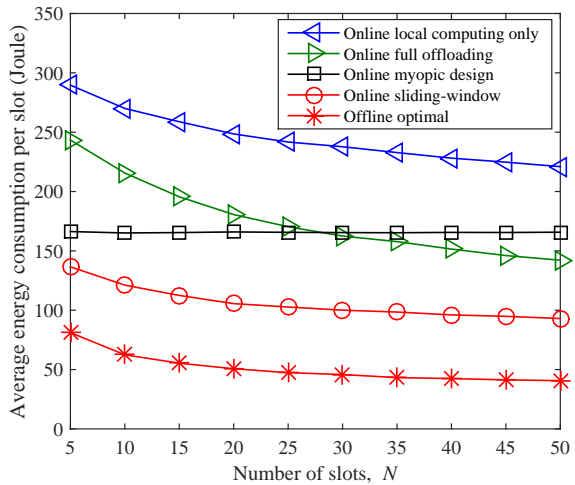}
 \caption{The average energy consumption per slot at the AP versus the number $N$ of slots.} \label{fig.vsN-online}
\end{figure}

Fig.~\ref{fig.vsN-online} shows the average energy consumption per slot at the AP versus the slot number $N$ of the horizon, where $\tau=0.02$ sec, $K=8$, $d_k=6$~m, and $A_{k,i}\in{\cal U}[0,8]$ Mbits, $\forall k\in{\cal K}$, $i\in{\cal N}$. The standard variances of the TSI/CSI prediction errors are set to be $\sigma_A=\sigma_h=\sigma_g=0.2$, and the sliding-window size is set to be $M=2$. It is observed in Fig.~\ref{fig.vsN-online} that the proposed online sliding-window based scheme outperforms the other three online schemes. As expected, due to imperfect prediction of future TSI/CSI and selection of the sliding-window size $M$, there exists a considerable performance gap between the proposed online scheme and the optimal offline one. Except for the online myopic-design scheme which optimizes joint-WPT-MEC design only based on the current slot's TSI/CSI, the average per-slot energy consumption achieved by the other schemes is observed to decrease as $N$ increases, similarly as Fig.~\ref{fig.vsN}. This illustrates the energy-saving benefit from exploiting the future (noisy) TSI/CSI. Furthermore, the online local-computing-only scheme is observed to perform inferior to the online full-offloading and myopic-design schemes. Finally, the online myopic-design scheme outperforms the online full-offloading one at small $N$ values (e.g., $N\leq 25$), but it is not true as $N$ grows larger in this setup.

\begin{figure}
  \centering
  \includegraphics[width = 3.3in]{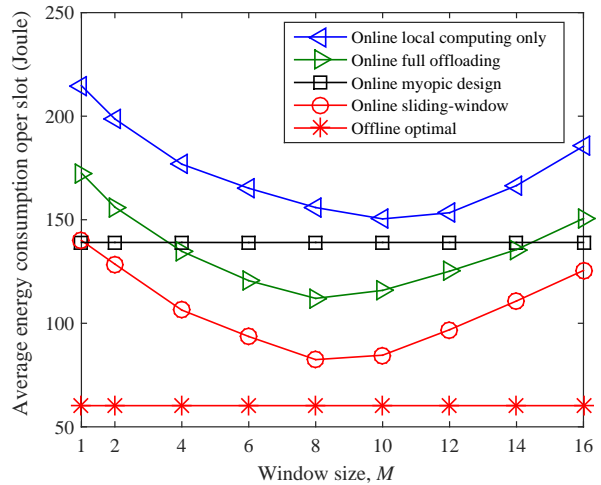}
 \caption{The average energy consumption per slot at the AP versus the window size $M$.} \label{fig.vsM}
\end{figure}

Fig. \ref{fig.vsM} shows the average energy consumption versus the sliding-window size $M$ for the proposed sliding-window based online scheme, where $\sigma_A=\sigma_h=\sigma_g=0.2$, $K=8$, $N=30$, $\tau=0.05$ sec, $d_k=5$ m, and $A_{k,i}\sim {\cal U}[1,5]$~Mbits, $\forall k\in{\cal K}$, $i\in{\cal N}$. It is observed that the proposed online sliding-window based scheme outperforms the other three online schemes. As expected, the performance gap between the proposed online scheme and the optimal offline scheme depends on the accuracy of the TSI/CSI prediction and the selection of the sliding-window size $M$. As $M$ increases, it is observed that the energy consumption achieved by the proposed online sliding-window scheme first decreases and then increases. On one hand, a large $M$ value can help exploit more knowledge of TSI/CSI in future to save the long-term system energy consumption in small predication-error cases. On the other hand, the performance benefits brought by large $M$ is more likely to be compromised by inaccuracy of the predicted TSI/CSI as $M$ increases beyond a certain value. As expected, there exists a tradeoff between choosing a large sliding-window size and keeping the accumulated error caused by imperfect TSI/CSI as small as possible. The similar performance trends are also observed for the online local-computing-only and full-offloading schemes. The online myopic design scheme outperforms the online local-computing-only scheme, but performs inferior to the online full-offloading scheme with proper selection of the sliding-window size. This further illustrates the benefit of exploiting the predicated TSI/CSI in online wireless powered MEC designs.

\begin{figure}
  \centering
  \includegraphics[width = 3.5in]{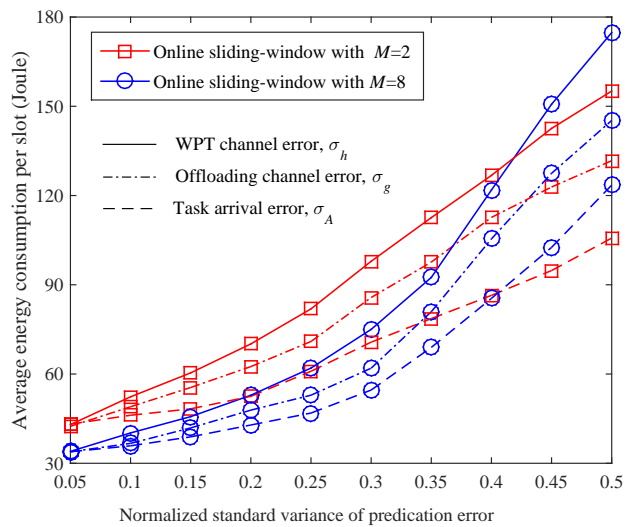}
 \caption{The average energy consumption per slot at the AP versus the predication errors.} \label{fig.vs_sigma}
\end{figure}

Fig.~\ref{fig.vs_sigma} shows the average energy consumption versus standard variances of the prediction errors $\sigma_A$, $\sigma_g$, and $\sigma_h$, respectively, where $N=20$, $K=4$, $\tau=0.1$ sec, $d_k=3$ m, and $A_{k,i}\sim{\cal U}[1,4]$~Mbits, $\forall k\in{\cal K}$, $i\in{\cal N}$. We set the standard variances of the TSI/CSI prediction errors as $\sigma_A=\sigma_g=\sigma_h=0.1$. It is observed in Fig.~\ref{fig.vs_sigma} that the energy consumption of the proposed sliding-window based online scheme increases as each of $\{\sigma_A,\sigma_g,\sigma_h\}$ increases. This is intuitively expected, since a large TSI (or CSI) prediction error incurs large energy consumption to compensate the TSI (or CSI) mismatch. It is also observed in Fig.~\ref{fig.vs_sigma} that the proposed online scheme with $M=8$ outperforms that with $M=2$ at small predication error values for each of $\{\sigma_A,\sigma_h,\sigma_g\}$, but it is not true at large predication error cases. This illustrates the importance to adapt the sliding-window size to the predication errors, in order to tradeoff the computational complexity for the performance of the proposed online scheme. It is also observed that the energy consumption due to the offloading CSI predication error $\sigma_g$ is larger than that due to the TSI predication error $\sigma_A$, since the former error additionally incurs transmission energy consumption due to the offloading channel mismatch. Finally, among the three types of predication errors, it is observed that the predication error of CSI for WPT $\sigma_h$ incurs the largest per-slot energy consumption with the same sliding-window size in this setup. This is because the users' energy supply relies solely on the AP's energy beamforming, which thus requires a sufficiently precise CSI predication for energy-efficient WPT.

\section{Concluding Remarks}
This paper studied the real-time online resource allocation scheme for a wireless powered multiuser MEC system to minimize the total system energy consumption over a finite horizon subject to the individual energy harvesting and task arrival causality constraints at users, by jointly optimizing the energy beamforming and remote task execution at the AP, as well as the local computing and task offloading strategies at the users over time. Using the Lagrange duality method, we obtained the optimal offline solution with the TSI/CSI perfectly known {\em a-priori}. We revealed that the optimal number of task input-bits allocated for the users' local computing and the AP's remote task execution should both monotonically increase over time. In the cases when the TSI/CSI are only causally known with their future information predicated with additive error, inspired by the optimal offline solution, we further proposed a sliding-window based online resource allocation scheme leveraging a sequential optimization. Numerical results demonstrated the merits of our proposed joint-WPT-MEC designs as compared to the other benchmark schemes without such joint design or with myopic based resource allocation only. It is envisioned that our results will provide a new approach for optimally integrating WPT and multiuser resource allocation in wireless powered dynamic MEC systems.

The current work may motivate several interesting research directions in the future, which are discussed as follows.
\begin{itemize}
  \item
  First, although this paper considers a FDMA based protocol for multiuser offloading in the uplink, the extension to time-division multiple access (TDMA) or non-orthogonal multiple access (NOMA) based offloading cases can be further pursued based on the approach developed in this paper. However, they require extra efforts made on top of the current joint-WPT-MEC designs, where equal-bandwidth allocation among the users is assumed throughout the whole time horizon. For example, in TDMA based offloading cases, multiuser scheduling along with the access time for each user needs to be designed within each slot, whilst in NOMA based offloading cases, the joint design of multiple users' offloading rates and the successive interference cancellation (SIC) decoding order at the AP should be addressed\cite{TCOM19}.
  \item
  Next, it is important to consider the long-term joint-WPT-MEC optimization for wireless powered MEC systems under more complex task dependency models (by considering the intra-user and inter-task dependency), in which the intra-user and inter-task dependency should be properly characterized\cite{Bi_new2020}.
  \item
  In addition, it is also interesting to consider the scenario when different computation deadline constraints may be imposed on different types of tasks. For this scenario, our results are generally extendable to this general scenario, by replacing the users' common deadline constraints in (10d) as several individual latency constraints for different tasks at different users. Since these individual latency constraints do not change the convexity of the energy minimization problem of interest, the similar methods used for solving problem (P1) in both online and offline cases are still applicable for solving the new problem. Nevertheless, the optimal offline solution structure (e.g., the monotonically increasing task allocation property for each user's local computing and the AP's remote computing) may not hold any more. It is challenging to obtain well-structured offline solutions in this scenario, which are left for our future work. Instead of finding the optimal offline solution, one potential solution is to simplify this problem by dividing the horizon into several sub-horizons according to the tasks' computation deadlines, and then apply the proposed solution in the single-common-deadline scenario to obtain a high-quality (though sub-optimal in general) solution.
  \item
   Finally, note that this paper focused on the latency-constrained energy minimization problem for delay-sensitive applications. In other scenarios with delay-tolerant (e.g., best-effort) applications, the computation rate, which is referred to as the average number of task input-bits executed per unit time, under communication/computation resource constraints may be a more relevant performance measure. In addition, by taking into account the system's energy consumption in computing/circuit/cooling and the operational expenses of underlying infrastructures, how to maximize the computation energy efficiency for wireless powered MEC systems is also an interesting problem worth further investigation.
  \end{itemize}

\appendix
\subsection{Proof of Theorem 1}
Let $\lambda_{k,i}\geq 0$, $\mu_{k,j}\geq 0$, $\mu_{k,N}\in\mathbb{R}$, $\nu_j \geq 0$, and $\nu_N\in\mathbb{R}$ denote the Lagrange multipliers associated with the constraints in (\ref{eq.prob1}b), (\ref{eq.prob1}c), (\ref{eq.prob1}d), (\ref{eq.prob1}e), and (\ref{eq.prob1}f), $j\in{\cal N}\setminus\{N\}$, $i\in{\cal N}$, $k\in{\cal K}$, respectively. The partial Lagrangian of problem (P1) is then given by
\begin{align}\label{eq.Lagrangian}
& {\cal L}(\{\lambda_{k,i},\mu_{k,i},\nu_i\},\{\bm S_i,l^{\rm mec}_{0,i},l^{\rm loc}_{k,i},l^{\rm off}_{k,i}\}) =  \notag \\
&~ \sum_{i=1}^{N} \tau {\rm tr}\Big(\bm S_i \breve{\bm H}_i\Big)+\sum_{i=1}^N \Big( \frac{\zeta_0 C_0^3 (l^{\rm mec}_{0,i})^3}{\tau^2} + \big( \sum_{j=i}^N\nu_j \big) l^{\rm mec}_{0,i} \Big)\notag\\
& +\sum_{k=1}^K \sum_{i=1}^N \Big(\frac{\left(\sum_{j=i}^N\lambda_{k,j}\right)\zeta_kC_k^3(l^{\rm loc}_{k,i})^3}{\tau^2}+\big( \sum_{j=i}^N\mu_{k,j}\big) l^{\rm loc}_{k,i}\Big) \notag \\ & -\sum_{k=1}^K\sum_{i=1}^{N} \big( \sum_{j=i}^N\mu_{k,j}\big) A_{k,i} + \sum_{k=1}^K\sum_{i=1}^{N} \Big[ \frac{\left(\sum_{j=i}^N\lambda_{k,j}\right) \tau\sigma^2}{\|{\bm g}_{k,i}\|^2}\times \notag\\
& \quad \Big(2^{\frac{l^{\rm off}_{k,i}}{\tau B}}-1 \Big) + \big( \sum_{j=i}^N(\mu_{k,j}-\nu_j)\big)l^{\rm off}_{k,i}\Big],
\end{align}
where $\breve{\bm H}_i\triangleq \bm I_{N_t} - \sum_{k=1}^K \left(\sum_{j=i}^N\lambda_{k,j}\right)\eta_k\bm h_{k,i}\bm h_{k,i}^H$, $\forall i\in{\cal N}$. Accordingly, the dual function of problem (P1) is expressed as
\begin{align}\label{eq.dual-func}
&{\cal G}( \{\lambda_{k,i},\mu_{k,i},\nu_i\} ) \triangleq \notag \\
&\min_{\{\bm S_i, l^{\rm mec}_{0,i},l^{\rm loc}_{k,i},l^{\rm off}_{k,i}\}} {\cal L}(\{\lambda_{k,i},\mu_{k,i},\nu_i\},\{\bm S_i,l^{\rm mec}_{0,i},l^{\rm loc}_{k,i},l^{\rm off}_{k,i}\}) \notag \\
&\quad {\rm s.t.}\quad \bm S_i\succeq \bm 0, l^{\rm mec}_{0,i}\geq 0, l^{\rm loc}_{k,i}\geq 0, l^{\rm off}_{k,i}\geq 0,~\forall i\in{\cal N}, k\in{\cal K}.
\end{align}
In order to obtain the dual problem of problem (P1), we first establish the following lemma.
\begin{lemma}\label{lem.bound}
In order for the dual function ${\cal G}(\{\lambda_{k,i},\mu_{k,i},\nu_i\})$ to be lower bounded from below, it must hold that
\begin{align}\label{eq.bound}
&\breve{\bm H}_i \succeq \bm 0,~\forall i\in{\cal N}, ~~\sum_{j=i}^N\lambda_{k,j} >0,~~\forall i\in{\cal N}, k\in{\cal K}.
\end{align}
\end{lemma}

\begin{IEEEproof}
First, we prove $\breve{\bm H}_i \succeq \bm 0$, $\forall i\in{\cal N}$. Suppose that $\breve{\bm H}_i$ is not semi-definite positive for some $i\in{\cal N}$. In this case, by letting $\bm S_i=\tau \bm x \bm x^H$ and $\tau \rightarrow \infty $, where $ \bm x^H \breve{\bm H}_i \bm x <0$ and $\bm x\in\mathbb{C}^{N_t\times 1}$, it can be shown from \eqref{eq.Lagrangian} that ${\cal L}(\cdot)\rightarrow -\infty$. Thus, it must hold that $\breve{\bm H}_i\succeq \bm 0$ for all $i\in{\cal N}$ in order for the dual function ${\cal G}(\{\lambda_{k,i},\mu_{k,i},\nu_{i}\})$ to be bounded from below. Next, we prove $\sum_{j=i}^N\lambda_{k,j} >0$, $\forall i\in{\cal N}, k\in{\cal K}$. Suppose that $\sum_{j=i}^N\lambda_j=0$ for some $k\in{\cal K}$ and $i\in{\cal N}$. In this case, by letting $\sum_{j=i}^N\mu_{k,j}<0$, $l^{\rm off}_{k,i}=0$, and $l^{\rm loc}_{k,i}\rightarrow \infty$, it can be shown from \eqref{eq.Lagrangian} that ${\cal L}(\cdot)\rightarrow -\infty$. Thus, the fact $\sum_{j=i}^N\lambda_{k,j}=0$ cannot be true in order to guarantee the dual function ${\cal G}(\{\lambda_{k,i},\mu_{k,i},\nu_{i}\})$ to be bounded from below. Lemma~\ref{lem.bound} is now proved.
\end{IEEEproof}

Based on Lemma \ref{lem.bound}, the dual problem of problem (P1) is then expressed as
\begin{subequations}\label{eq.prob1_dual}
\begin{align}
({\rm D}1):~&\max_{\{\lambda_{k,i},\mu_{k,i},\nu_i\} } ~{\cal G}(\{\lambda_{k,i},\mu_{k,i},\nu_i\})\\
&~~~~~~{\rm s.t.} ~~(\ref{eq.bound}),~ \lambda_{k,i} \geq 0, \mu_{k,j}\geq 0, \nu_j \geq 0, \notag\\
&~~~~~~~~~~~~~~~~\forall i\in{\cal N},j\in{\cal N}\setminus\{N\},k\in{\cal K}.
\end{align}
\end{subequations}
Denote $\cal X$ as the feasible solution set of $\{\lambda_{k,i},\mu_{k,i},\nu_i\}$ for problem \eqref{eq.prob1_dual}. Then, we solve problem (P1) optimally by solving its dual problem (D1).

Under any given $\{\lambda_{k,i},\mu_{k,i},\nu_i\}\in{\cal X}$, by removing the irrelevant constant terms in \eqref{eq.Lagrangian}, the optimization problem in \eqref{eq.dual-func} can be equivalently decomposed into the $(2N+2NK)$ independent subproblems for different time slots and users, each of which is expressed as follows for one $i$ or $(k,i)$ with $k\in{\cal K}$ and $i\in{\cal N}$:
\begin{align}
&\min_{\bm S_i} ~~{\rm tr}\left(\bm S_i \breve{\bm H}_i\right)~~\quad{\rm s.t.}~\bm S_i\succeq \bm 0 \label{eq.subp1}\\
&\min_{l^{\rm mec}_{0,i}}~ \frac{\zeta_0 C_0^3 (l^{\rm mec}_{0,i})^3}{\tau^2} + \Big(\sum_{j=i}^N\nu_j\Big) l^{\rm mec}_{0,i},~~{\rm s.t.}~l^{\rm mec}_{0,i}\geq 0 \label{eq.subp1-2}\\
&\min_{l^{\rm loc}_{k,i}} \frac{\left(\sum_{j=i}^N\lambda_{k,j}\right)\zeta_k C_k^3(l^{\rm loc}_{k,i})^3}{\tau^2}+\Big( \sum_{j=i}^N\mu_{k,j}\Big)l^{\rm loc}_{k,i} \notag\\
 &~{\rm s.t.}~l^{\rm loc}_{k,i}\geq 0 \label{eq.subp2}\\
&\min_{l^{\rm off}_{k,i}} \frac{\left(\sum_{j=i}^N\lambda_{k,j}\right) \tau \sigma^2 (2^{\frac{l^{\rm off}_{k,i}}{\tau B}}-1)} {\|{\bm g}_{k,i}\|^2}+ \Big(\sum_{j=i}^N(\mu_{k,j}-\nu_j) \Big) l^{\rm off}_{k,i}\notag \\
&~~{\rm s.t.}~l^{\rm off}_{k,i}\geq 0. \label{eq.subp3}
\end{align}
We denote $\bm S_i^\prime$, $l^{\rm mec \prime}_{0,i}$, $l^{\rm loc \prime}_{k,i}$, and $l^{\rm off \prime}_{k,i}$ as the optimal solutions to the subproblems in \eqref{eq.subp1}, \eqref{eq.subp1-2}, \eqref{eq.subp2}, and \eqref{eq.subp3}, respectively. We then establish the following lemmas.
\begin{lemma}\label{lem.sub1}
The optimal solution $\bm S_i^\prime$ to \eqref{eq.subp1} is given by $\bm S_i^\prime \in {\rm Null}({\breve{\bm H}})$, $i\in{\cal N}$, where ${\rm Null}(\cdot)$ denotes the null space of a matrix.
\end{lemma}
\begin{IEEEproof}
As the matrix $\breve{\bm H}_{i}$ is semidefinite positive in problem \eqref{eq.subp1}, with $\bm S_i\succeq \bm 0$ , it follows that the minimal value of ${\rm tr}(\bm S_i\breve{\bm H}_{i})$ is zero. Therefore, the optimal $\{\bm S_i^\prime\}$ to problem \eqref{eq.subp1} must satisfy $\bm S_i^\prime\in{\rm Null}(\breve{\bm H}_{i})$, $\forall i\in{\cal N}$.
\end{IEEEproof}

\begin{lemma}\label{lem.sub1-2}
 The optimal solution $l^{\rm mec \prime}_{0,i}$ to \eqref{eq.subp1-2} is given by
$l^{\rm mec \prime}_{0,i} = \frac{\tau}{\sqrt{3}\zeta^{\frac{1}{2}}_0C_0^{\frac{3}{2}}}\sqrt{\big( -\sum_{j=i}^N\nu_j\big)^+}$.
\end{lemma}
\begin{IEEEproof}
Introducing the Lagrange multiplier $\theta \geq 0$ associated with the constraint $l^{\rm mec}_{0,i}\geq 0$, the Lagrangian of problem \eqref{eq.subp1-2} is given as
${\cal L}_0(l^{\rm mec}_{0,i},\theta) = \frac{\zeta_0 C_0^3 (l^{\rm mec}_{0,i})^3}{\tau^2} + \left(\sum_{j=i}^N\nu^*_j\right) l^{\rm mec}_{0,i} - \theta l^{\rm mec}_{0,i}$. Accordingly, the dual problem of problem \eqref{eq.subp1-2} is expressed as
\begin{align}\label{eq.subp1-2-dual}
\max_{\theta\geq 0}\min_{l^{\rm mec}_{0,i}\geq 0}{\cal L}_0(l^{\rm mec}_{0,i},\theta)
\end{align}
Note that problem \eqref{eq.subp1-2} is a convex optimization problem and satisfies Slater's condition\cite{Boyd2004}. Therefore, strong duality holds between problems \eqref{eq.subp1-2} and \eqref{eq.subp1-2-dual}. Let $l_{0,i}^{\rm mec\prime}$ and $\theta^\prime$ be the optimal solutions for problems \eqref{eq.subp1-2} and \eqref{eq.subp1-2-dual}, respectively. The Karush-Kuhn-Tucker (KKT) optimal conditions are then given by\cite{Boyd2004}
\begin{subequations}\label{eq.subp1-2-dual-KKT}
\begin{align}
&\frac{2\zeta_0C_0^3(l_{0,i}^{\rm mec\prime})^2}{\tau^2}+\sum_{j=i}^N \nu_j -\theta^\prime = 0\\
&\theta^\prime l^{\rm mec\prime}_{0,i}=0,~~l^{\rm mec\prime}_{0,i}\geq 0,~~\theta^\prime \geq 0,
\end{align}
\end{subequations}
where (\ref{eq.subp1-2-dual-KKT}a) denotes that the gradient of ${\cal L}_0(l^{\rm mec}_{0,i},\theta)$ must vanish at $l_{0,i}^{\rm mec\prime}$ and the first equality in (\ref{eq.subp1-2-dual-KKT}b) is the complementary slackness condition. Since $\theta^\prime$ acts as a slack variable in (\ref{eq.subp1-2-dual-KKT}a), it can then be eliminated. Together with $l^{\rm mec\prime}_{0,i}\geq 0$, we obtain $l^{\rm mec \prime}_{0,i}=
\frac{\tau}{\sqrt{3}\zeta^{\frac{1}{2}}_0C_0^{\frac{3}{2}}}\sqrt{\left( -\sum_{j=i}^N\nu_j\right)^+}$.
\end{IEEEproof}

\begin{lemma}\label{lem.sub2}
 The optimal solution $l^{\rm loc \prime}_{k,i}$ to \eqref{eq.subp2} is given by
$l_{k,i}^{\rm loc \prime} = \frac{\tau}{\sqrt{3}\zeta_k^{\frac{1}{2}}C_k^{\frac{3}{2}}} \sqrt{\frac{\big(-\sum_{j=i}^N \mu_{k,j}\big)^+}{\sum_{j=i}^N \lambda_{k,j}}}$.
\end{lemma}

\begin{lemma}\label{lem.sub3}
 The optimal solution $l^{\rm off \prime}_{k,i}$ to \eqref{eq.subp3} is given by
\begin{align}\label{eq.dual-sub3}
&l_{k,i}^{\rm off \prime} = \tau B\log_2\times\notag\\
&~~\left(\max\left[1,~ \frac{\sum_{j=i}^N\big(\nu_{j}-\mu_{k,j}\big)}{\sum_{j=i}^N\lambda_{k,j}}
\Big/\left(\frac{\tau\sigma^2\ln2}{B\|\bm g_{k,i}\|^2}\right)\right]\right).
\end{align}
\end{lemma}

Note that Lemmas \ref{lem.sub2} and \ref{lem.sub3} can be are similarly proved as for Lemma \ref{lem.sub1-2}, and thus we omit the proofs for brevity. Based on Lemma \ref{lem.sub1}, it follows that the optimal transmit covariance matrix $\bm S_i^\prime$ for WPT is not unique if $\breve{\bm H}_i$ is rank deficient (i.e., ${\rm rank}(\breve{\bm H}_i) < N_t$). As a result, we can simply choose $\bm S_i^\prime=\bm 0$ as an optimal solution to problem \eqref{eq.subp1} for all $i\in{\cal N}$ in the procedure of evaluating the dual function. With Lemmas 3--5 and $\bm S_i^\prime=\bm 0$, $\forall i\in{\cal N}$, one can readily evaluate the dual function ${\cal G}(\{\lambda_{k,i},\mu_{k,i},\nu_i\})$ under given $\{\lambda_{k,i},\mu_{k,i},\nu_i\}\in{\cal X}$.

Next, we maximize the dual function ${\cal G}(\{\lambda_{k,i},\mu_{k,i},\nu_i\})$ over $\{\lambda_{k,i},\mu_{k,i},\nu_i\}\in{\cal X}$ to solve the dual problem (D1). Note that the dual function ${\cal G}(\{\lambda_{k,i},\mu_{k,i},\nu_i\})$ is always concave but not necessarily differentiable. Therefore, problem (D1) is convex and can thus be solved by subgradient based methods such as the ellipsoid method \cite{Boyd2004}.

With the dual optimal $\{\lambda^{*}_{k,i},\mu^{*}_{k,i},\nu^{*}_i\}$ obtained, it remains to find the optimal primal solution to problem \eqref{eq.prob1}. For each $i\in{\cal N}$ and $k\in{\cal K}$, since $l^{\rm mec\prime}_{0,i}$, $l_{k,i}^{\rm loc\prime}$ and $l_{k,i}^{\rm off\prime}$ are the uniquely optimal solutions to problems \eqref{eq.subp1-2}, \eqref{eq.subp2}, and \eqref{eq.subp3}, respectively, the optimal $l_{0,i}^{\rm mec*}$, $l_{k,i}^{\rm loc*}$, and $l_{k,i}^{\rm off*}$ to problem~(P1) can be directly obtained by replacing $\lambda_{k,i}$, $\mu_{k,i}$, $\nu_i$ with the optimal dual $\lambda^{*}_{k,i}$, $\mu^{*}_{k,i}$, $\nu^{*}_i$ in Lemmas 3--5, respectively. Therefore, we obtain the optimal solution of $\{l_{0,i}^{\rm mec*},l_{k,i}^{\rm loc*},l_{k,i}^{\rm off*}\}$ to problem~(P1), as given by (\ref{eq.optimal-sol}a)--(\ref{eq.optimal-sol}c).

On the other hand, since the dual optimal $\bm S_i^\prime=\bm 0$ of problem~\eqref{eq.subp1}, $\forall i\in{\cal N}$, are even not feasible for problem \eqref{eq.prob1}, the optimal $\bm S_i^{*}$ of problem (P1) cannot be obtained from Lemma~\ref{lem.sub1} alone. Therefore, an additional procedure is required to obtain the optimal $\{\bm S_i^{*}\}$ to problem~(P1). Specifically, with $\{l_{0,i}^{\rm mec*},l_{k,i}^{\rm loc*},l_{k,i}^{\rm off*}\}$ obtained already, the optimal $\{\bm S_i^{*}\}$ to problem~(P1) is obtained by solving the following convex semidefinite program (SDP) via off-the-shelf convex solvers (e.g., CVX toolbox \cite{cvx}):
\begin{subequations}\label{eq.prob-Q}
\begin{align*}
 & \{\bm S_i^{*}\}\triangleq \argmin_{\{\bm S_i\succeq \bm 0\}} ~\sum_{i=1}^{N} \tau {\rm tr}(\bm S_i) \\
&\quad\quad\quad\quad {\rm s.t.} ~ \sum_{j=1}^i\Big( \frac{\zeta_kC^3_k(l^{\rm loc*}_{k,j})^3}{\tau^2} + \frac{\tau \sigma^2(2^{\frac{l^{\rm off*}_{k,j}}{\tau B}}-1)}{\|{\bm g}_{k,j}\|^2} \Big) \notag \\
&\quad\quad\quad\quad\quad\quad \leq \sum_{j=1}^{i} \tau\eta_k{\rm tr}(\bm S_j \bm h_{k,j} \bm h^H_{k,j}),~ \forall k\in{\cal K}, i\in{\cal N}.
\end{align*}
\end{subequations}
By combining $\{\bm S_i^{*}\}$ together with $\{ l_{0,i}^{\rm mec*}, l_{k,i}^{\rm loc*},l_{k,i}^{\rm off*}\}$, we finally obtain the optimal solution to problem \eqref{eq.prob1}. Until now, we complete the proof of Theorem 1.

\subsection{Proof of Proposition \ref{prop1}}
First, we consider the number of $l_{0,i}$ task input-bits for the AP's remote task execution. Based on Lemma~\ref{lem.sub1-2}, $\forall i\in{\cal N}\setminus\{N\}$, it follows that $
l_{0,i+1}^{\rm mec*} = \sqrt{\frac{N\tau^2 \left(-\sum_{j=i+1}^N\nu^{*}_{j}\right)^+}{3\zeta_0 C_0^3}} \geq \sqrt{\frac{N\tau^2 \left(-\sum_{j=i}^N\nu^{*}_{j}\right)^+}{3\zeta_0 C_0^3}} = l_{0,i}^{\rm mec*}$, where the inequality follows from the fact that $\nu^{*}_{i}\geq 0$ for any $i\in{\cal N}\setminus\{N\}$. It thus holds that $0=l^{\rm mec*}_{0,1}\leq l^{\rm mec*}_{0,2}\leq \ldots \leq l_{0,N}^{\rm mec*}$.

Next, we consider the case with $k\in{\cal K}$. Similarly, from Lemma~\ref{lem.sub2}, $\forall i\in{\cal N}\setminus\{N\}$, it yields that $l^{\rm loc*}_{k,i+1} = \sqrt{\frac{\tau^2\left(-\sum_{j=i+1}^N \mu_{k,j}^{*}\right)^+}{3\sum_{j=i+1}^N \lambda^{*}_{k,j}\zeta_kC^3_k}}\geq \sqrt{\frac{\tau^2\left(-\sum_{j=i}^N \mu_{k,j}^{*}\right)^+}{3\sum_{j=i}^N \lambda^{*}_{k,j}\zeta_kC^3_k}} = l^{\rm loc*}_{k,i}$, $\forall k\in{\cal K}$, where the inequality follows from the fact that both $\mu_{k,i}^{*}\geq 0$ and $\lambda_{k,i}^{*}\geq 0$ hold for any $i\in{\cal N}\setminus\{N\}$. Therefore, we have $l^{\rm loc*}_{k,1}\leq\ldots \leq l_{k,N}^{\rm loc*}$ for all $k\in{\cal K}$.

\subsection{Proof of Proposition \ref{Proposition-myopic}}
For the myopic design scheme, due to the task causality and completion constraints, it follows that $\hat{l}^{\rm mec}_{0,1}=0$ and $\hat{l}^{\rm off}_{k,N}=0$, $\forall k\in{\cal N}$. Then, we consider the the myopic design scheme at slot $i\in{\cal N}\setminus\{N\}$, in which problem (P1) can be reduced into
\begin{subequations}\label{eq.prob1-myopic}
\begin{align}
&\min_{\bm S_i,\{l^{\rm loc}_{k,i},l^{\rm off}_{k,i}\} } \tau{\rm tr}(\bm S_i) + \frac{\zeta_0 C_0^3 (l^{\rm mec}_{0,i+1})^3}{\tau^2}\\
&~~\quad {\rm s.t.}~ \frac{\zeta_k C_k^3 (l^{\rm loc}_{k,i})^3}{\tau^2} + \frac{\tau\sigma^2}{\|{\bm g}_{k,i}\|^2}(2^{\frac{l^{\rm off}_{k,i}}{\tau B}}-1) \notag \\
& \quad\quad\quad\quad\quad\quad \leq  \tau \eta_k {\rm tr}\left(\bm S_i\bm h_{k,i}\bm h_{k,i}^H \right),\quad \forall k \in{\cal K}\\
&~~ \quad\quad\quad A_{k,i} - l^{\rm loc}_{k,i} - l^{\rm off}_{k,i} = 0, \quad \forall k \in{\cal K}\\
&~~ \quad\quad\quad \bm S_i\succeq \bm 0, l^{\rm loc}_{k,i}\geq 0,l^{\rm off}_{k,i}\geq 0,~~ \forall k\in{\cal K},
\end{align}
\end{subequations}
where $ l^{\rm mec}_{0,i+1}- \sum_{k=1}^K l^{\rm off}_{k,i} = 0$, $\forall i\in{\cal N}\setminus\{N\}$. By substituting $l^{\rm loc}_{k,i}=A_{k,i}-l^{\rm off}_{k,i}$ into \eqref{eq.prob1-myopic}, we obtain the following univariate optimization problem:
\begin{align}\label{eq.prob1-myopic1}
&\min_{0\leq l^{\rm off}_{k,i} \leq A_{k,i}} \frac{\zeta_k C_k^3 (A_{k,i}-l^{\rm off}_{k,i})^3}{\tau^2} + \frac{\tau\sigma^2}{\|{\bm g}_{k,i}\|^2}(2^{\frac{l^{\rm off}_{k,i}}{\tau B}}-1).
\end{align}
Since the cubic function $x^3$ and exponential function $2^x$ with respect to $x\geq 0$ are both convex, it is ready to show that \eqref{eq.prob1-myopic1} is a convex optimization problem\cite{Boyd2004}. Therefore, based on the first-order derivative condition of \eqref{eq.prob1-myopic1}, the optimal number of task input-bits for user $k$'s offloading $\hat{l}_{k,i}^{\rm off}$ is obtained by solving the univariate equation of $\frac{3\zeta_kC_k^3(A_{k,i}-l_{k,i}^{\rm off})^2}{\tau^2} = \frac{\tau \sigma^2 \ln2}{\|\bm g_{k,i}\|^2} 2^{\frac{l_{k,i}^{\rm off}}{\tau B}}$ at slot $i\in{\cal N}\setminus\{N\}$.

With $\hat{l}_{k,i}^{\rm off}$ at hand, the optimal number of task input-bits for user $k$'s local computing $\hat{l}_{k,i}^{\rm loc}$ and AP's remote task execution $\hat{l}_{0,i}^{\rm mec}$ are obtained in (\ref{Proposition-myopic-sol}a) and (\ref{Proposition-myopic-sol}b), respectively. Furthermore, by substituting the obtained $\hat{l}_{k,i}^{\rm off}$ and $\hat{l}_{k,i}^{\rm loc}$ into \eqref{eq.prob1-myopic}, the optimal covariance matrices $\hat{\bm S}_i$'s for the AP's energy beamforming are obtained in (\ref{Proposition-myopic-sol}a). Until now, Proposition~\ref{Proposition-myopic} is proved.

\end{document}